\documentclass{optica-article}

\journal{opticajournal} 

\articletype{Research Article}

\usepackage{lineno}

\usepackage{amsmath}
\usepackage{subcaption}
\usepackage{xr}

\newcommand{\etalcite}[2]{#1 \textit{et al.}~\cite{#2}}

\begin{document}

\title{Coupling loss and transmission in a multimode fiber-fed Virtually Imaged Phased Array (VIPA)}

\author{Matthew C. H. Leung,\authormark{1,2,*} Andrew Szentgyorgyi,\authormark{1,2,3} \\Colby Jurgenson,\authormark{1,3} and David Charbonneau\authormark{1,2}}

\address{\authormark{1}Center for Astrophysics | Harvard \& Smithsonian, 60 Garden Street, Cambridge, MA 02138, USA\\
\authormark{2}Department of Astronomy, Harvard University, 60 Garden Street, Cambridge, MA 02138, USA\\
\authormark{3}Smithsonian Astrophysical Observatory, 100 Acorn Park Drive, Cambridge, MA 02140, USA}

\email{\authormark{*}matthew.leung@cfa.harvard.edu} 


\begin{abstract*} 
The Virtually Imaged Phased Array (VIPA) is a spectral disperser that has seen increasing adoption across various applications, including optical telecommunications, high-resolution spectroscopy, and LiDAR. Although VIPAs are typically fed by single-mode optical fiber in these applications, there is growing interest in using a multimode optical fiber feed to enable higher throughput. However, a multimode fiber feed introduces several challenges, one of which is a fundamental limit on how much light can be coupled into a VIPA. In this work, we derive a closed-form expression for the minimum tilt angle of a multimode fiber-fed VIPA required for no etendue coupling loss, and we show that this angle depends solely on the input fiber's etendue, the focal ratio of the injection cylindrical lens, and the VIPA's intrinsic parameters. Moreover, we derive the 2D and 3D etendue that can be coupled into a VIPA, as well as the VIPA's etendue coupling efficiency and total transmission, given a particular configuration. We validate our mathematical models using ray-tracing simulations in Zemax OpticStudio non-sequential mode. Our results provide practical guidance for the design of high-throughput optical systems based on multimode fiber-fed VIPAs.

\end{abstract*}



\section{Introduction}

The Virtually Imaged Phased Array (VIPA) is a relatively new spectral disperser based on a Fabry-Perot etalon, offering several advantages over conventional dispersers like a diffraction grating. Compared to a grating, VIPAs provide higher angular dispersion, a more compact form factor, and low polarization-dependence \cite{Shirasaki1996}, while having higher transmission compared to a conventional Fabry-Perot etalon \cite{Weiner2012}. VIPAs were initially developed for wavelength division multiplexing \cite{Shirasaki1996,Xiao2005} and dispersion compensation \cite{Shirasaki1997} in optical telecommunications, but their use has since expanded to a broad range of fields. Applications involving VIPAs fed by optical fiber have been growing, such as spectrally encoded imaging \cite{Metz2014}, light detection and ranging (LiDAR) \cite{Li2021}, and high-resolution spectroscopy, which includes applications like molecular spectroscopy \cite{NugentGlandorf2012}, Brillouin spectroscopy \cite{Scarcelli2011,Bouvet2024}, and astronomy \cite{Bourdarot2018,Zhu2020,Carlotti2022,Zhu2023,Leung2025}.

In such applications, VIPAs are typically fed by single-mode optical fiber rather than multimode optical fiber, because of the single-mode fiber's more stable beam profile, its well-defined spatial and angular extent that can be approximated by a Gaussian beam \cite{SalehTeich}, and its small etendue \cite{Bourdarot2018}, which generally makes optical design easier. However, there is recent growing interest in using a multimode fiber feed to allow for more light to pass through the optical system \cite{Zhu2023,Bouvet2024,Leung2025}, as it is easier to couple light into a multimode fiber compared to the smaller single-mode fiber. The use of a multimode fiber, however, introduces several challenges, one of which is a limit to how much light can be coupled into a VIPA. All optical systems are fundamentally limited by the conservation of etendue \cite{nonimagingoptics_chaves}, and the VIPA is no exception, with there being a maximum etendue that can be coupled into a particular VIPA \cite{Bouvet2024}.

To date, there has been no comprehensive study examining how the physical parameters of a multimode fiber-fed VIPA system affect the constraints required to minimize coupling loss, and how these same parameters affect VIPA transmission. One important parameter that affects coupling loss is the VIPA's tilt angle; if the tilt angle is too shallow, then not all of the light can stay inside the VIPA. Most analyses in the literature assume that a Gaussian beam is coupled into the VIPA, as is appropriate for a single-mode fiber feed. For example, \etalcite{Xiao}{Xiao2005} derived a constraint on the minimum VIPA tilt angle for no coupling loss under the assumption of a Gaussian beam input, which is not a suitable model for a multimode fiber feed. In addition, existing models of VIPA transmission typically assume an infinitely-long VIPA and depend only on the reflectances of the VIPA's surfaces \cite{Weiner2012}, without accounting for the finite VIPA length or the etendue losses that arise with a multimode fiber feed.

In this work, we develop a mathematical framework for coupling light into a VIPA using geometrical optics. We derive a constraint for the minimum tilt angle of a multimode fiber-fed VIPA required for no etendue coupling loss, expressions for the two-dimensional~(2D) and three-dimensional~(3D) etendue that can be coupled into a particular VIPA, and the VIPA transmission. We explicitly relate each of these quantities to experimentally-relevant physical parameters in the optical system used to couple light into a VIPA. These results provide practical design rules for high-throughput optical systems that use multimode fiber-fed VIPAs. This is of particular relevance to light-starved applications such as astronomical spectroscopy of faint and distant objects.

Our paper is organized as follows. In Section~\ref{sec:problem_setup}, we define the geometry and coordinate conventions used throughout our derivations. In Section \ref{sec:tilt_angle}, we derive a closed-form expression for the minimum VIPA tilt angle for no etendue coupling loss. In Section~\ref{sec:etendue_coupled}, we derive the 2D and 3D etendue that can be coupled into a VIPA, and hence the VIPA etendue coupling efficiency. In Section~\ref{sec:transmission}, we derive the transmitted etendue and the VIPA transmission. In Section~\ref{sec:Zemax_sim}, we validate our mathematical models using ray-tracing simulations in Ansys Zemax OpticStudio non-sequential mode. Readers are invited to consult Table \ref{tab:reserved_symbols} throughout the paper for a summary of selected variables used in this work.

\begin{table}[htbp]
\caption{Table of selected reserved symbols}
  \label{tab:reserved_symbols}
  \centering
\begin{tabular}{ccc}
\hline
Symbol & Variable & Reference \\
\hline
$a$ & Radius of optical fiber & -- \\
$F$ & Focal ratio of light coming out of optical fiber & -- \\
$h$ & VIPA thickness & -- \\
$n'$ & Refractive index of VIPA & -- \\
$n$ & Refractive index of surrounding medium & -- \\
$f_\mathrm{coll}$ & Focal length of collimator & -- \\
$f_c$ & Focal length of cylindrical lens & -- \\
$F_c$ & Focal ratio of cylindrical lens & -- \\
$D_c$ & Diameter of cylindrical lens & -- \\
$\beta$ & VIPA tilt angle & -- \\
$\tilde{\beta}$ & Minimum VIPA tilt angle for no etendue coupling loss & Eq.~(\ref{eq:beta_min})\\
$\theta_F$ & Cone half-angle corresponding to $F_c$ & Eq.~(\ref{eq:theta_F}) \\
$N$ & VIPA normal coordinate & Eq.~(\ref{eq:N})\\
$T$ & VIPA tangential coordinate & Eq.~(\ref{eq:T})\\
$d$ & Distance between cylindrical lens and VIPA for focus & Eq.~(\ref{eq:d})\\
$T_{0r2}$ & Tangential coordinate of ray 2 at $N=0$ & Eq.~(\ref{eq:T0r2}) \\
$G_{\mathrm{full}}$ & 3D etendue of optical fiber & Eq.~(\ref{eq:G_full}) \\
$G_{\mathrm{3D}}$ & 3D etendue coupled into VIPA & Eq.~(\ref{eq:3D_etendue_solution}) \\
$R_1$ & VIPA HR surface reflectance & -- \\
$R_2$ & VIPA PR surface reflectance & -- \\
$L$ & VIPA HR surface length & -- \\
\hline
\end{tabular}
\end{table}


\section{Problem Setup}\label{sec:problem_setup}

In this section, we explain the optical system for coupling light into the VIPA, and we set up the necessary geometry conventions for derivations in subsequent sections.

\subsection{VIPA overview and optical system for coupling fiber light into a VIPA}

\begin{figure}[htbp]
\centering\includegraphics[width=0.7\textwidth]{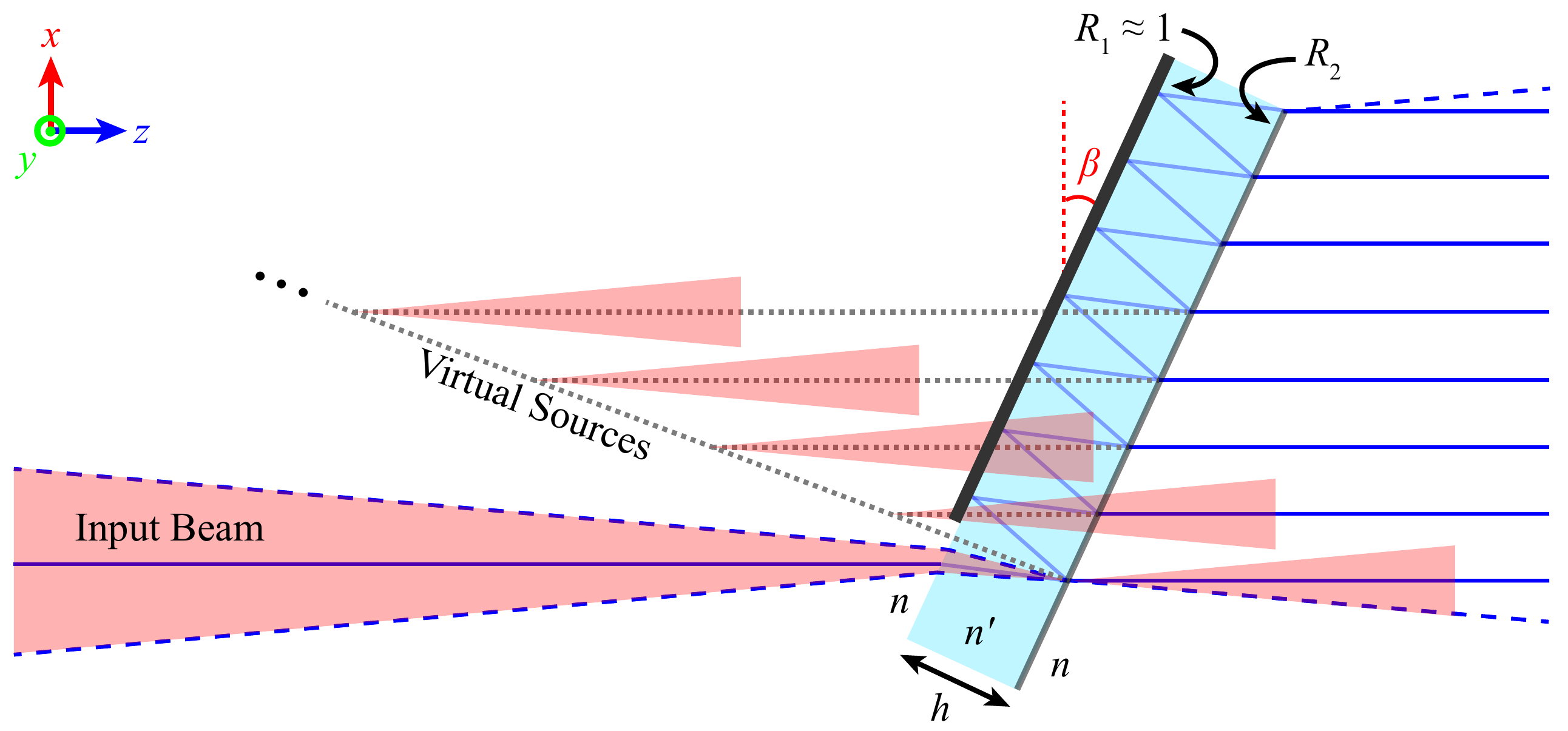}
\caption{Illustration of a Virtually Imaged Phased Array (VIPA). Multiple reflections between the two internally reflective surfaces create a phased array of virtual sources that interfere to achieve high spectral resolution.}\label{fig:VIPA}
\end{figure}

Fig.~\ref{fig:VIPA} shows an illustration of a VIPA. The VIPA is essentially a modified Fabry-Perot etalon with a transmissive entrance window. The back surface has an internally partially-reflective (PR) coating. The front surface, excluding the transmissive entrance window, has an internally high-reflective (HR) coating with reflectance near 1. The transmissive entrance window has an anti-reflection coating. Like in a Fabry-Perot etalon, the beam put into the VIPA accumulates an optical phase difference on every round trip between the two internally reflective surfaces. When this round-trip phase difference is an integer multiple of $2\pi$, a resonance condition is satisfied, resulting in an intensity peak. Since the round-trip phase difference depends on wavelength, changing the wavelength changes the output angle at which the resonance condition is satisfied. As a result, the VIPA is a spectral disperser, dispersing light along the $x$-direction in Fig.~\ref{fig:VIPA}.

\begin{figure}[htbp]
\centering\includegraphics[width=\textwidth]{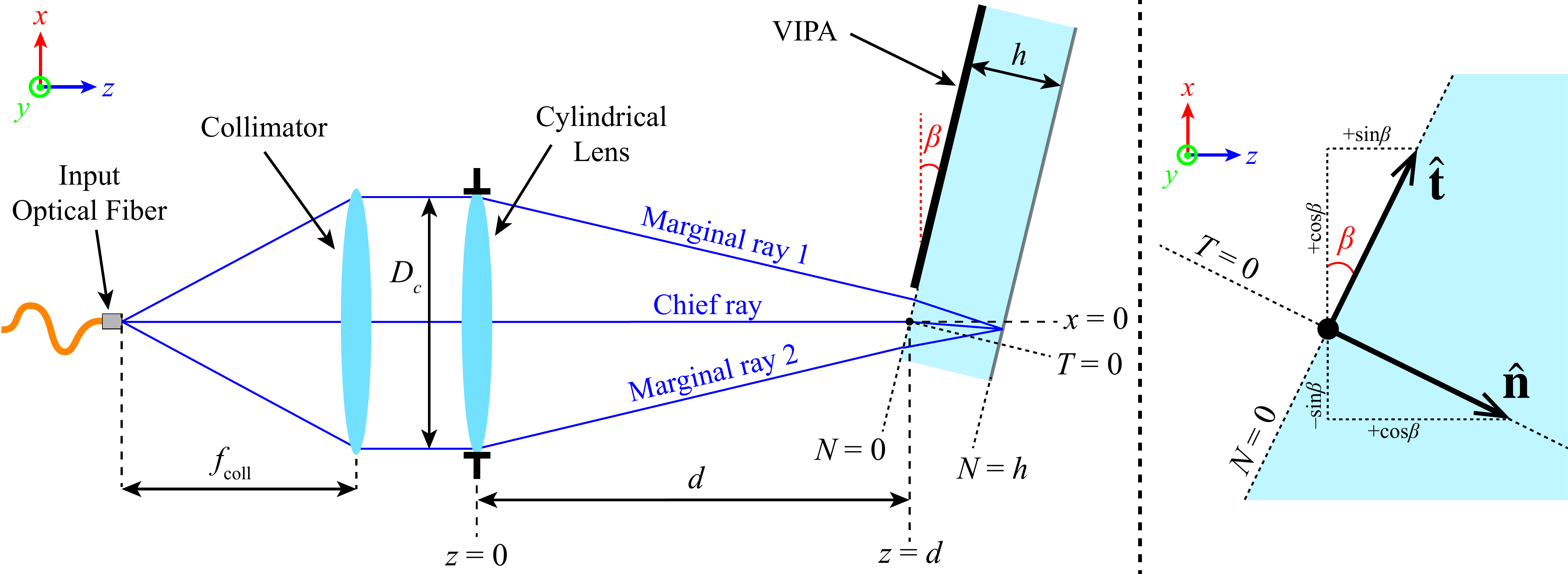}
\caption{Optical system for coupling light from an optical fiber into the VIPA. This optical system is comprised of a collimator with focal length $f_\mathrm{coll}$ and a cylindrical lens with focal length $f_c$. We take the stop of the optical system to be the cylindrical lens, which has diameter $D_c$. The $z$-axis is the optical axis. $d$ is the distance, along the optical axis, between the cylindrical lens and the VIPA front surface, such that the beam forms focus on the VIPA back surface. The VIPA front and back surfaces are at the normal coordinates $N=0$ and $N=h$ respectively. The chief ray intersects the VIPA front surface at the tangential coordinate $T=0$.}\label{fig:VIPA_coordinate_system}
\end{figure}

Fig.~\ref{fig:VIPA_coordinate_system} shows a typical optical system for coupling light from an optical fiber into a VIPA. We assume that the optical fiber has a circular core with radius $a$, and that light is coming out of the fiber at a focal ratio $F$, in a cone whose half-angle is $\arctan{(1/(2F))}$. Light coming out of the fiber is collimated by a collimator with focal length $f_\mathrm{coll}$. We assume that the focal ratio of the collimator is matched to $F$.

The collimated beam is then line-focused in the $x$-direction by a cylindrical lens with focal length $f_c$ onto the VIPA back surface. We take the cylindrical lens to be the stop of the optical system. The cylindrical lens has diameter $D_c$ and focal ratio $F_c=f_c/D_c$.

\subsection{Problem geometry}

We will now define the geometry of the optical system, as illustrated in Figs.~\ref{fig:VIPA} and \ref{fig:VIPA_coordinate_system}. Let the $z$-axis be the optical axis. Fig.~\ref{fig:VIPA_coordinate_system} shows three blue rays that originate from the on-axis object point at the fiber, corresponding to object height $x=0$. The chief ray from this on-axis field point coincides with the optical axis. Consider the VIPA in Fig.~\ref{fig:VIPA}, with thickness $h$ and refractive index $n'$, in a surrounding medium with refractive index $n$. Let $R_1$ and $R_2$ be the reflectances of the front and back surfaces of the VIPA respectively. We use the same geometry as \etalcite{Hu}{Hu2015}, where the symmetry plane of the VIPA is located in the $xz$ plane. Let the VIPA tilt angle be $\beta$ in the $xz$ plane. Let $d$ be the distance, along the $z$-axis, between the cylindrical lens and VIPA front surface, such that the beam forms focus on the VIPA back surface as illustrated in Fig.~\ref{fig:VIPA_coordinate_system}.

The cylindrical lens is situated at $z=0$, and the chief ray intersects the VIPA front surface at $z=d$. The VIPA front surface is the plane $z=d+x\tan{\beta}$. For convenience, let us define some VIPA-frame-centric coordinates. The VIPA inward normal unit vector $\hat{\mathbf{n}}$ and tangential unit vector are $\hat{\mathbf{t}}$ are:
\begin{subequations}
\begin{align}
    \hat{\mathbf{n}} &= (-\sin{\beta})\hat{\mathbf{x}}+(\cos{\beta})\hat{\mathbf{z}} = (-\sin{\beta}, \cos{\beta})\\
    \hat{\mathbf{t}} &= (\cos{\beta})\hat{\mathbf{x}}+(\sin{\beta})\hat{\mathbf{z}} = (\cos{\beta}, \sin{\beta})
\end{align}
\end{subequations}
as illustrated in Fig.~\ref{fig:VIPA_coordinate_system}. Then we can define the VIPA-frame-centric coordinates $(N,T)$ by:
\begin{subequations}
\begin{align}
    N &\equiv \hat{\mathbf{n}} \cdot ((x,z)-(0,d)) = (-\sin{\beta})x + (\cos{\beta})(z-d) \label{eq:N}\\
    T &\equiv \hat{\mathbf{t}} \cdot ((x,z)-(0,d)) = (\cos{\beta})x + (\sin{\beta})(z-d) \label{eq:T}
\end{align}
\end{subequations}
We call $N$ the normal coordinate and $T$ the tangential coordinate. Then the VIPA front and back surfaces are at $N=0$ and $N=h$ respectively. The chief ray intersects the VIPA front surface at $(N,T)=(0,0)$.


\subsection{Ray tracing}\label{sec:ray_tracing}

\begin{figure}[htbp]
\centering\includegraphics[width=\textwidth]{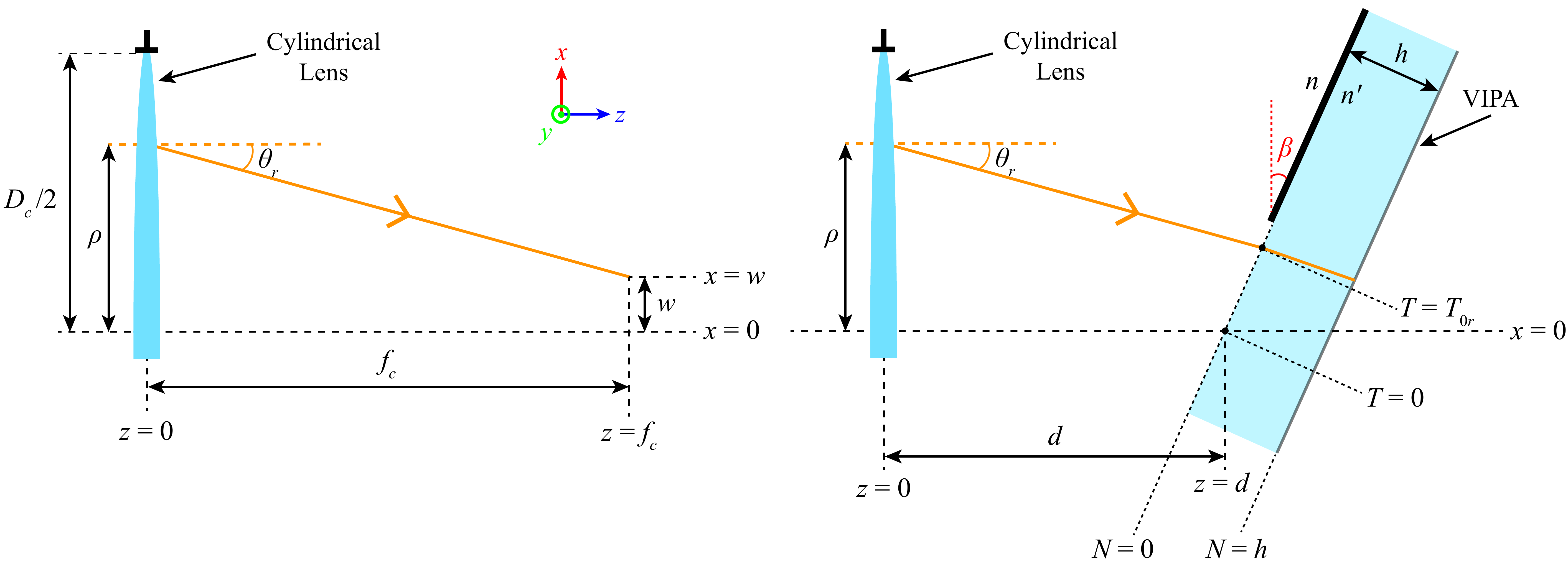}
\caption{The path of an arbitrary ray with height $\rho$ at $z=0$ and angle $\theta_r$ with respect to the $z$-axis. In the absence of the VIPA, the ray would have a height $w$ at $z=f_c$. With the VIPA present, the ray intersects the VIPA front surface at tangential coordinate $T=T_{0r}$, which is given by Eq.~(\ref{eq:T_0r}). The tangential coordinate of this ray, as a function of normal coordinate $N$, is given by Eq.~(\ref{eq:T_r}).}\label{fig:VIPA_arbitrary_ray}
\end{figure}

Our derivations in the subsequent sections rely on knowledge of the tangential coordinates of certain rays, at particular normal coordinates. Consider an arbitrary ray shown in orange in Fig.~\ref{fig:VIPA_arbitrary_ray}. This ray has a height of $\rho$ at the cylindrical lens (at $z=0$), where $\rho\in[-D_c/2,D_c/2]$. This ray makes an angle $\theta_r$ with respect to the $z$-axis. In the absence of a VIPA, this ray would have a height $w$ at the cylindrical lens' focal plane ($z=f_c$). Our convention is that the particular $\rho$, $\theta_r$, and $w$ in Fig.~\ref{fig:VIPA_arbitrary_ray} are all positive. With some algebra, we can find the tangential coordinate $T_r(N)$ of this ray at an arbitrary normal coordinate $N$:
\begin{subequations}
\begin{align}
    T_r(N) &= 
    \begin{cases} 
      \displaystyle T_{0r} + N\tan{\theta_r'} & \mathrm{if}\:\:\: N\geq0 \\
      \displaystyle T_{0r} + N\tan{(\beta-\theta_r)} & \mathrm{if}\:\:\: N<0
   \end{cases} \label{eq:T_r}\\
   \theta_r&=\arctan{\left(\frac{\rho-w}{f_c}\right)} \label{eq:theta_r} \\
   \theta_r'&=\arcsin{\left(\frac{n}{n'}\sin{(\beta-\theta_r)}\right)} \label{eq:thetap_r} \\
   T_{0r} &= \frac{(f_c-d)\sin{\theta_r}+w\cos{\theta_r}}{\cos{(\beta-\theta_r)}} \label{eq:T_0r}
\end{align}
\end{subequations}
Formally, $N<0$ and $N>h$ is when the ray is outside of the VIPA, and $0\leq N \leq h$ is when the ray is inside the VIPA. However, Eq.~(\ref{eq:T_r}) can also be used to describe the ray's tangential coordinate after reflection from the internally reflective surfaces. If a ray remains inside the VIPA, we can use Eq.~(\ref{eq:T_r}) to determine its tangential coordinate with an appropriate choice of $N$, where $N\geq 0$. For example, after the first reflection off the VIPA back surface, the ray still inside the VIPA would have a tangential coordinate corresponding to $h<N<2h$. After the next reflection off the VIPA front surface, the ray still inside the VIPA would have a tangential coordinate corresponding to $2h<N<3h$. We will use this fact in derivations in the subsequent sections.


\subsection{Distance for focus at VIPA back surface}

One quantity of relevance in our work is $d$, which, recall, is the distance between the cylindrical lens and VIPA front surface, such that the beam forms focus on the VIPA back surface. We take this to be when marginal ray 1 and the chief ray intersect at $N=h$, as illustrated in Fig.~\ref{fig:VIPA_coordinate_system}. The result is:
\begin{subequations}\label{eq:d}
\begin{align}
    d &= f_c-\frac{h\cos{(\theta_F-\beta)}}{\sin{\theta_F}} (\tan{\theta_{cr}'} - \tan{\theta_{m1}'}) \label{eq:d_only}\\
    \theta_{m1}' &= \arcsin{\left(\frac{n}{n'}\sin{(\beta-\theta_F)}\right)} \label{eq:thetap_m1}\\
    \theta_{cr}' &= \arcsin{\left(\frac{n}{n'}\sin{\beta}\right)} \label{eq:thetap_cr}\\
    \theta_F &\equiv \arctan{\left(\frac{1}{2F_c}\right)} \label{eq:theta_F}
\end{align}
\end{subequations}
A full derivation can be found in Supplement 1. Here, $\theta_{m1}'$ and $\theta_{cr}'$ are the angles that marginal ray 1 and the chief ray respectively make with the VIPA normal, inside of the VIPA. $\theta_F$ is the cone half-angle corresponding to the focal ratio of the cylindrical lens $F_c$.


\section{Minimum VIPA tilt angle for no coupling loss}\label{sec:tilt_angle}

In this section, we derive the minimum VIPA tilt angle for no etendue coupling loss, given $a/F$, $F_c$, $h$, and $n/n'$. \etalcite{Xiao}{Xiao2005} had derived the minimum VIPA tilt angle for a Gaussian beam input to the VIPA, which is appropriate for a single-mode fiber feed. However, a multimode fiber has a spatial and angular extent not adequately described by a Gaussian beam model. We instead derive the minimum VIPA tilt angle using a geometrical optics treatment based on the fiber etendue.

\subsection{Geometry for maximum etendue coupling}\label{sec:tilt_angle_geometry}

\begin{figure}[htbp]
\centering\includegraphics[width=0.8\textwidth]{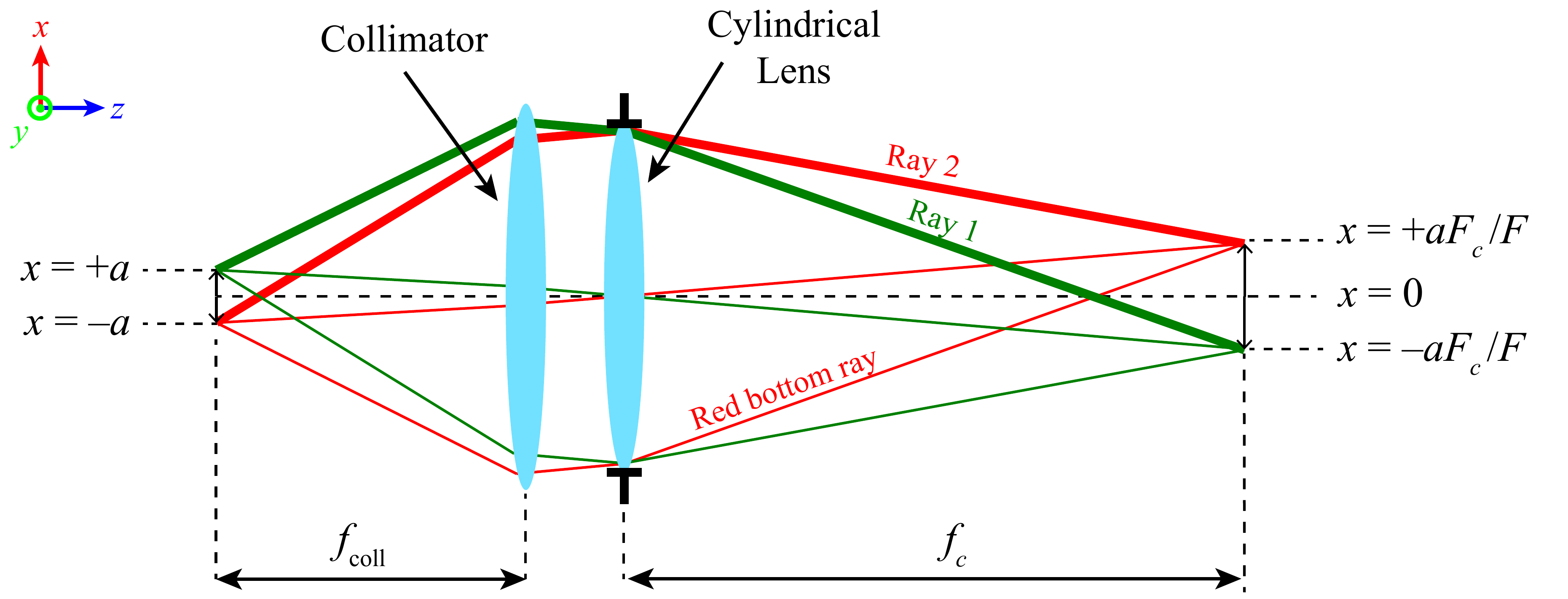}
\caption{Illustration of two ray bundles originating from object heights $x=a$ (green) and $x=-a$ (red) at the fiber, corresponding to the spatial extent of the fiber. Of particular interest are the two marginal rays labeled ``Ray 1'' and ``Ray 2''.}\label{fig:VIPA_2lens_ray1ray2}
\end{figure}

Let us first consider the geometry for the largest possible etendue coupled into a VIPA. Consider Fig.~\ref{fig:VIPA_2lens_ray1ray2}, which shows two ray bundles. The green and red ray bundles originate from two off-axis object points at the fiber, corresponding to object heights $x=a$ and $x=-a$ respectively. These are object heights corresponding to the spatial extent of the fiber. In the absence of a VIPA, the green and red ray bundles will form focus at the cylindrical lens focal plane ($z=f_c$), at image heights of $x=-\frac{aF_c}{F}$ and $x=\frac{aF_c}{F}$ respectively. Of particular interest are the two marginal rays labeled ``Ray 1'' and ``Ray 2'' in Fig.~\ref{fig:VIPA_2lens_ray1ray2}.

\begin{figure}[htbp]
  \centering
  \begin{subfigure}{0.6\textwidth}
    \centering
    \includegraphics[width=\textwidth]{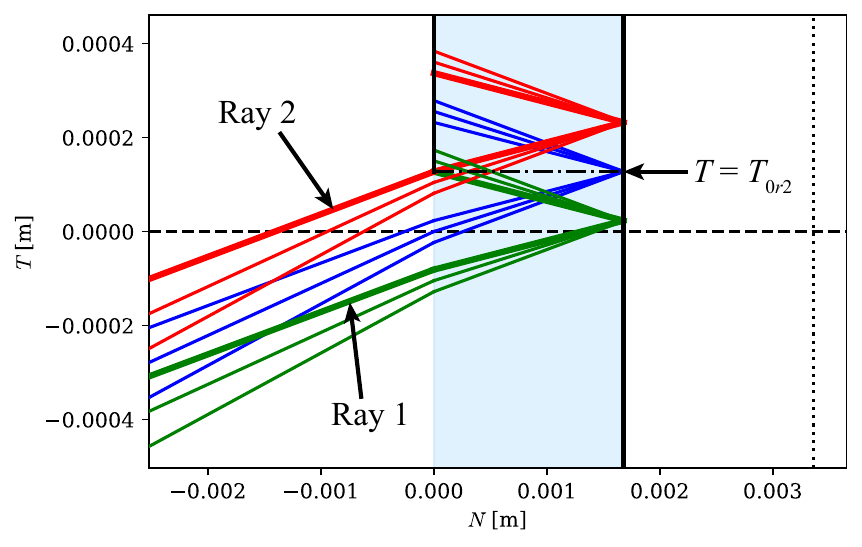}
    \caption{Trace of rays inside VIPA.}
    \label{fig:VIPA_ray_interior}
  \end{subfigure}
  \hfill
  \begin{subfigure}{0.6\textwidth}
    \centering
    \includegraphics[width=\textwidth]{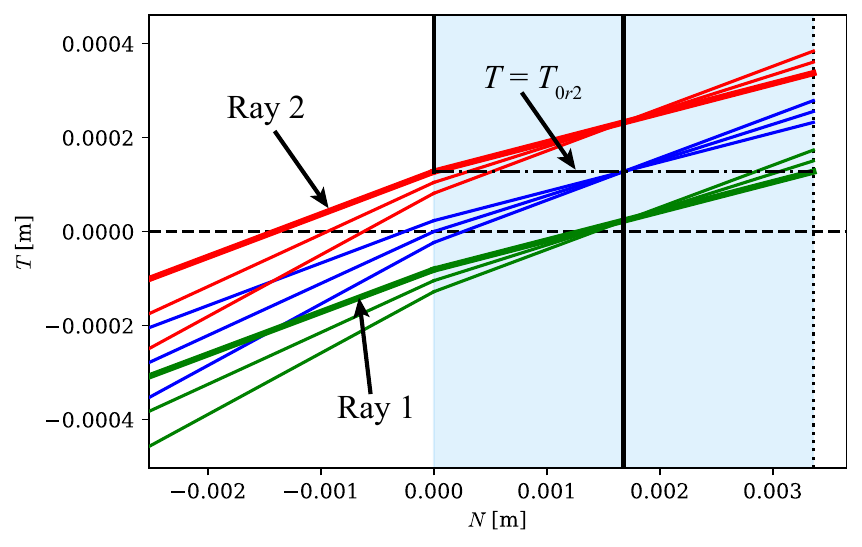}
    \caption{Expanded version of Fig.~\ref{fig:VIPA_ray_interior}, where $N>h$ refers to after the first reflection off the VIPA back surface.}
    \label{fig:VIPA_ray_expanded}
  \end{subfigure}
  \caption{Illustration of geometry for maximum etendue coupling into a VIPA. This is a ray trace. The green, blue, and red ray bundles originate from object heights $x=a$, $x=0$, and $x=-a$ respectively at the fiber. Here, we use $h=1.68$ mm, $n'=1.4494$ (fused silica), $n=1.0003$ (air), $a=25$ \textmu m, $F=6$, and $F_c=25$. Here, the VIPA is oriented at the minimum tilt angle for no etendue coupling loss, which, computed from Eq.~(\ref{eq:beta_min}), is $\tilde{\beta}=6.310^\circ$. The dotted-dashed line shows $T=T_{0r2}$, which is the tangential coordinate of ray~2 at the VIPA front surface ($N=0$). Notice that in this geometry, ray~1 intersects $T=T_{0r2}$ at $N=2h$.}
  \label{fig:VIPA_ray_input}
\end{figure}

In order for the largest possible etendue to be coupled into the VIPA, there are two conditions. The first condition is that the VIPA must be positioned such that the HR surface begins above the point where ray 2 first intersects the VIPA front surface, so that all of the incident light can pass through the transmissive entrance window. This is illustrated in Fig.~\ref{fig:VIPA_ray_input}. However, if the VIPA tilt angle is too shallow, then the light that entered the VIPA can also escape back through the entrance window after the first reflection off the back surface. Hence, the second condition is that the VIPA must be tilted such that after ray 1 makes its first reflection off the back surface, it will intersect the VIPA front surface at the point where ray 2 entered the VIPA. This geometry is illustrated in Fig.~\ref{fig:VIPA_ray_input}. Hence, for no etendue coupling loss, the minimum required VIPA tilt angle is one such that ray 2's tangential coordinate at $N=0$ is equal to ray 1's tangential coordinate at $N=2h$.

\subsection{Closed-form expression for minimum VIPA tilt angle for no coupling loss}

Let $T_{r1}(N)$ and $T_{r2}(N)$ be the tangential coordinates of ray 1 and ray 2 respectively, at some normal coordinate $N\geq 0$ inside the VIPA. $T_{r1}(N)$ and $T_{r2}(N)$ can be found from Eqs.~(\ref{eq:theta_r}), (\ref{eq:T_0r}), (\ref{eq:thetap_r}), and (\ref{eq:T_r}), using $\rho=D_c/2$ and $w=-\frac{aF_c}{F}$ for ray 1, and $\rho=D_c/2$ and $w=\frac{aF_c}{F}$ for ray 2. Then the mathematical condition for no etendue coupling loss is $T_{r1}(2h) = T_{r2}(0)$. In particular, let us define $T_{0r2}\equiv T_{r2}(0)$ as the tangential coordinate of ray 2 at the VIPA front surface. This $T_{0r2}$ quantity will be of particular importance in the subsequent sections. Let $\tilde{\beta}$ be the VIPA tilt angle such that $T_{r1}(2h) = T_{r2}(0)$. The solution for $\tilde{\beta}$ is: 
\begin{subequations}\label{eq:beta_min}
\begin{align}
    \tilde{\beta} &= \arcsin{\left(\sqrt{\frac{-c_2-\sqrt{c_2^2-4c_1c_3}}{2c_1}}\right)} + \arctan{\left(\frac{1}{2F_c}\right)} \label{eq:beta_min_only}\\
    c_1 &\equiv \left( \frac{hn}{n'}\right)^2 \label{eq:c_1}\\
    c_3 &\equiv \left(\frac{a}{F} \right)^2 \frac{4F_c^4}{4F_c^2+1} \label{eq:c_3} \\
    c_2 &\equiv -c_1 \left(\frac{c_3}{h^2} + 1 \right) \label{eq:c_2}
\end{align}
\end{subequations}
A full derivation can be found in Supplement 1. In the derivation, we have made the approximation $\frac{a}{f_\mathrm{coll}} \ll \frac{1}{2F_c}$, which is valid since the fiber radius $a$ is several orders of magnitude smaller than the collimator focal length $f_\mathrm{coll}$. Eq.~(\ref{eq:beta_min}) is the minimum VIPA tilt angle for no coupling loss. Notice how this depends only on 4 parameters: the VIPA thickness $h$, the ratio between the surrounding medium's refractive index and the VIPA's refractive index $n/n'$, the cylindrical lens focal ratio $F_c$, and $a/F$, which is a proxy for the input fiber's etendue (see the end of Section~\ref{sec:tilt_angle_analysis}). Also, notice how $\tilde{\beta}$ is independent of the collimator and cylindrical lens focal lengths, $f_\mathrm{coll}$ and $f_c$, and is independent of $d$. There is a notable dependence on the cylindrical lens focal ratio $F_c$.

\begin{figure}[htbp]
\centering\includegraphics[width=0.7\textwidth]{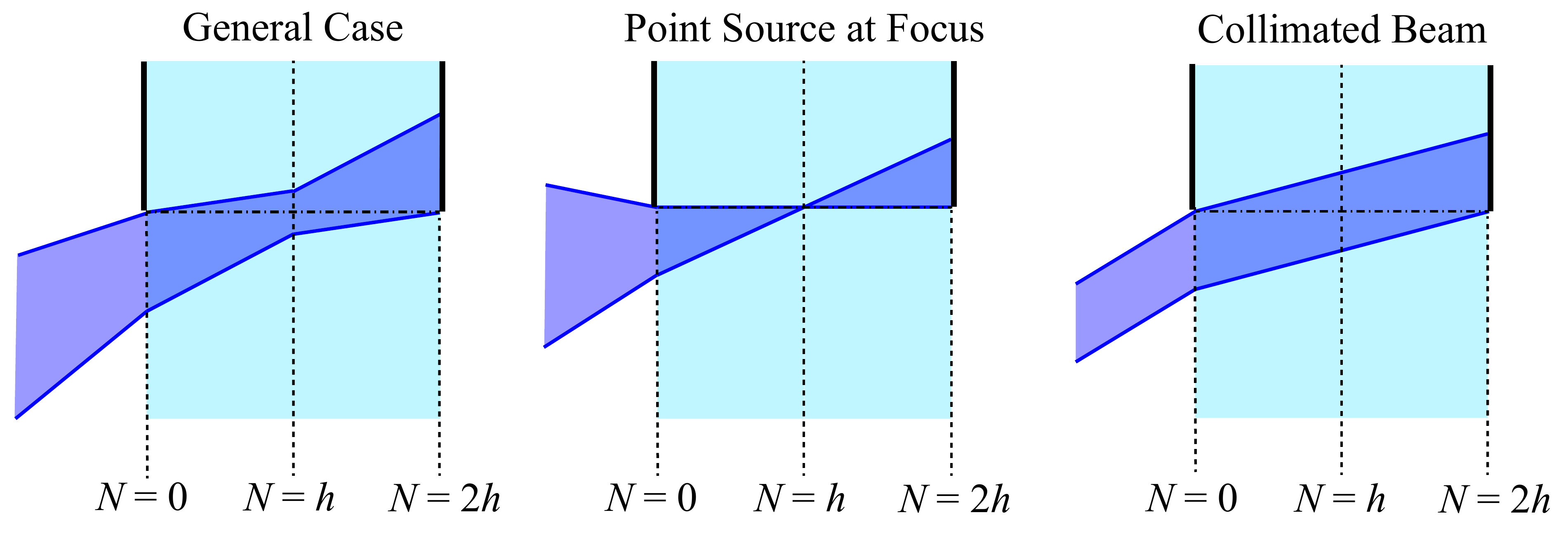}
\caption{Three cases for no etendue coupling loss, with three different kinds of input beams. The left illustration shows the general case, where $\tilde{\beta}$ is given by Eq.~(\ref{eq:beta_min}). The center illustration shows the case where there is a point source focus at the VIPA back surface ($N=h$), where $\tilde{\beta}$ is given by Eq.~(\ref{eq:beta_point_source}). The right illustration shows the case where a collimated beam is fed into the VIPA, where $\tilde{\beta}$ is given by Eq.~(\ref{eq:beta_min_coll}).}\label{fig:VIPA_feed_3scenarios}
\end{figure}

There are two special cases for Eq.~(\ref{eq:beta_min}), depending on $F_c$. When $F_c$ is very small, there is effectively a point source at focus. When $F_c$ is very large, effectively a collimated beam is being fed into the VIPA. These two cases will be discussed in Sections \ref{sec:beta_point_source} and \ref{sec:beta_collimated_beam} respectively. These two cases are illustrated in Fig.~\ref{fig:VIPA_feed_3scenarios}.

\subsection{Minimum VIPA tilt angle for a point source}\label{sec:beta_point_source}

\begin{figure}[htbp]
\centering\includegraphics[width=0.37\textwidth]{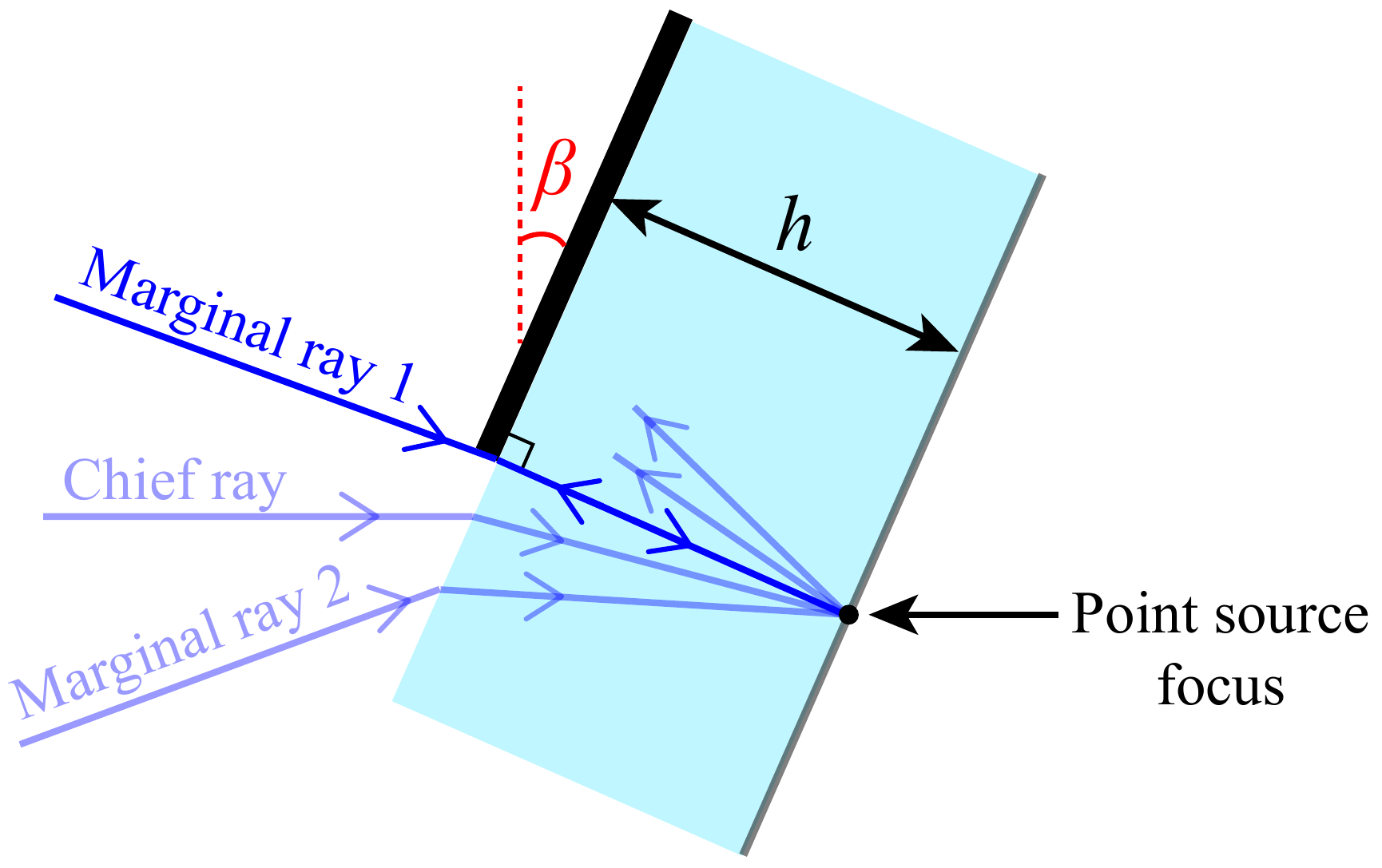}
\caption{Geometry for the minimum VIPA tilt angle for no etendue coupling loss, when there is a point source focus at the VIPA back surface.}\label{fig:VIPA_point_source_feed}
\end{figure}

When $F_c$ is small or when $a$ is small, we effectively have a point source focus at the back surface of the VIPA. In order for there to be no coupling loss, the minimum VIPA tilt angle $\tilde{\beta}$ must take on a value such that marginal ray 1 makes an angle $\theta_{m1}'=0$ with respect to the VIPA normal, as illustrated in Fig.~\ref{fig:VIPA_point_source_feed}. Using Eq.~(\ref{eq:thetap_m1}), this happens when $n\sin{(\tilde{\beta}-\theta_F)}=n'\sin{(0)}$, which implies that:
\begin{equation}\label{eq:beta_point_source}
    \tilde{\beta} = \theta_F = \arctan{\left(\frac{1}{2F_c}\right)}
\end{equation}
Hence, for a point source focus at the back surface of the VIPA, the the minimum VIPA tilt angle for no coupling loss is simply the cone half-angle $\theta_F$ corresponding to $F_c$. We can see this clearly if we substitute $a=0$ into Eq.~(\ref{eq:beta_min}). If $a=0$, then by Eq.~(\ref{eq:c_3}) we have $c_3=0$, which by Eq.~(\ref{eq:c_2}) implies that $c_2=-c_1$, which then implies that $-c_2-\sqrt{c_2^2-4c_1c_3}=0$. Thus, the $\arcsin$ term in Eq.~(\ref{eq:beta_min_only}) becomes 0, and hence we obtain Eq.~(\ref{eq:beta_point_source}). Eq.~(\ref{eq:beta_min}) reduces to Eq.~(\ref{eq:beta_point_source}) when $a=0$. 


\subsection{Minimum VIPA tilt angle for a collimated beam}\label{sec:beta_collimated_beam}

When $F_c$ is large, we effectively have a collimated beam entering the VIPA. If the incident beam diameter outside of the VIPA is $W$, then the minimum VIPA tilt angle $\tilde{\beta}$ for no coupling loss is:
\begin{equation}\label{eq:beta_min_coll}
    \tilde{\beta} = \arcsin{\left(\sqrt{\frac{n^2(4h^2+W^2)-\sqrt{n^4(4h^2+W^2)^2-(4Wh n n' )^2}}{8h^2 n^2}}\right)}
\end{equation}
A full derivation can be found in Supplement 1. Eq.~(\ref{eq:beta_min_coll}) leads to an interesting result, which is that as $W$ increases, $\tilde{\beta}$ also increases, but only up to a certain point. After this point, no solutions of $\tilde{\beta}$ are possible. The largest possible $\tilde{\beta}$ occurs when the discriminant term is zero, where we will have:
\begin{equation}\label{eq:max_beta_min}
    \max{(\tilde\beta)} = \arcsin{\left(\sqrt{ \left(\frac{n'}{n}\right)^2 - \frac{n'}{n}\sqrt{\left(\frac{n'}{n}\right)^2-1}} \right)}
\end{equation}
Eq.~(\ref{eq:max_beta_min}) shows that the largest possible minimum VIPA tilt angle for no coupling loss is set entirely by the ratio between the refractive indices of the VIPA and its surrounding medium. Eq.~(\ref{eq:max_beta_min}) is a good approximation for the largest possible $\tilde{\beta}$ in the general case Eq.~(\ref{eq:beta_min}).


\subsection{Analysis and Discussion}\label{sec:tilt_angle_analysis}

\begin{figure}[htbp]
\centering\includegraphics[width=0.6\textwidth]{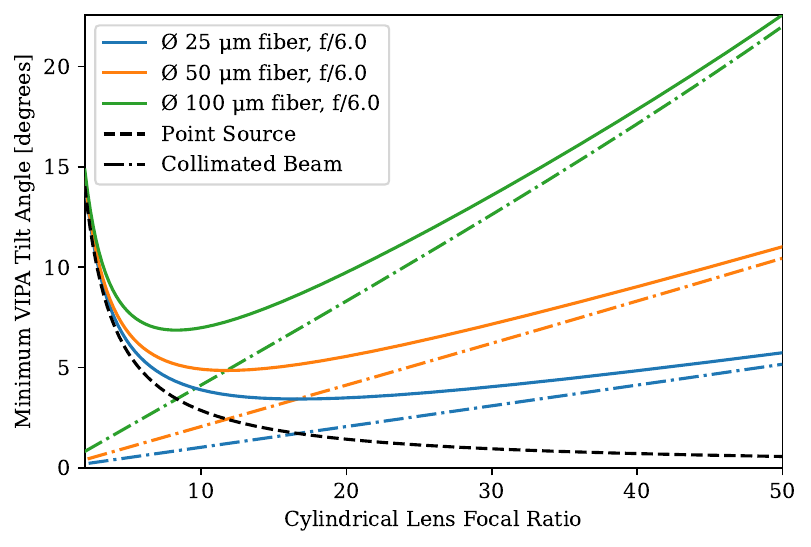}
\caption{Minimum VIPA tilt angle $\tilde{\beta}$ for no etendue coupling loss as a function of cylindrical lens focal ratio $F_c$, for different $a/F$ values. The blue, orange, and green sets of curves correspond to $a=12.5,25,50$~\textmu m respectively, with $F=6$. Here, we use $h=1.68$ mm, $n'=1.4494$ (fused silica), and $n=1.0003$ (air). The solid curves show Eq.~(\ref{eq:beta_min}). The dashed curve shows the minimum tilt angle for a point source focus at the VIPA back surface, given by Eq.~(\ref{eq:beta_point_source}). The dotted-dashed curves show the minimum tilt angle for a collimated beam input, given by Eq.~(\ref{eq:beta_min_coll}), assuming the relation given by Eq.~(\ref{eq:W_coll_beam_relation}). For small $F_c$ or small $a$, the solid curve approaches the dashed curve. For large $F_c$, the solid curve approaches the corresponding dotted-dashed curve.}\label{fig:tilt_angle}
\end{figure}

Fig.~\ref{fig:tilt_angle} shows $\tilde{\beta}$ as a function of $F_c$ for several different $a/F$ values. The solid curves show $\tilde{\beta}$ as computed using Eq.~(\ref{eq:beta_min}). The dashed curve shows $\tilde{\beta}$ for a point source focus on the VIPA back surface, as computed using Eq.~(\ref{eq:beta_point_source}). The dotted-dashed curves show $\tilde{\beta}$ for a collimated beam input to the VIPA, as computed using Eq.~(\ref{eq:beta_min_coll}). In order to have a meaningful comparison, for these dotted-dashed curves, we take the diameter of the collimated beam to be:
\begin{equation}\label{eq:W_coll_beam_relation}
    W = (2a)\frac{F_c}{F}
\end{equation}
This value of $W$ is the spatial extent of the fiber image at the cylindrical lens focal plane, in the absence of the VIPA (cf. Fig.~\ref{fig:VIPA_2lens_ray1ray2}).

In Fig.~\ref{fig:tilt_angle}, we see several interesting relationships. First, note that for a fixed $a/F$, $h$, and $n/n'$, the minimum tilt angle $\tilde{\beta}$ is convex in $F_c$. The function $\tilde{\beta}(F_c)$ has a global minimum at some $F_c$. Let $F_c^*$ be the cylindrical lens focal ratio such that $\tilde{\beta}(F_c^*)$ is the global minimum. If $F_c>F_c^*$, then $\tilde{\beta}$ is a monotonically increasing function of $F_c$. When $F_c$ is large, $\tilde{\beta}$ approaches the collimated beam result given by Eq.~(\ref{eq:beta_min_coll}). That is, each solid curve approaches its corresponding dotted-dashed curve in Fig.~\ref{fig:tilt_angle}.

If $F_c<F_c^*$, then $\tilde{\beta}$ is a monotonically decreasing function of $F_c$. When $F_c$ is small, $\tilde{\beta}$ approaches the point source result given by Eq.~(\ref{eq:beta_point_source}). That is, each solid curve approaches the dashed curve in Fig.~\ref{fig:tilt_angle}. When $a/F$ is small, $\tilde{\beta}$ also approaches the point source result. If a fiber has a larger $a/F$, then the required $\tilde{\beta}$ increases.

\begin{figure}[htbp]
\centering\includegraphics[width=0.65\textwidth]{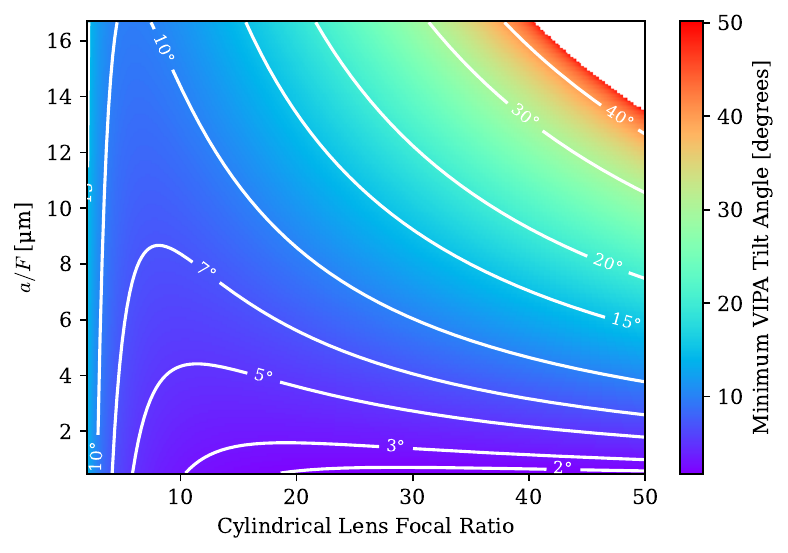}
\caption{Minimum VIPA tilt angle $\tilde{\beta}$ for no etendue coupling loss (Eq.~(\ref{eq:beta_min})) as a function of cylindrical lens focal ratio $F_c$ and the etendue proxy $a/F$. Here, we use $h=1.68$ mm, $n'=1.4494$ (fused silica), and $n=1.0003$ (air). The white region in the top right corner indicates combinations of $a/F$ and $F_c$ where no physical solutions of $\tilde{\beta}$ exist. This happens when roughly $\tilde{\beta} \gtrsim \max{\tilde{\beta}}=49.61^\circ$, given by Eq.~(\ref{eq:max_beta_min}).}\label{fig:tilt_angle_pcolormesh}
\end{figure}

Another way to understand the minimum VIPA tilt angle is to plot $\tilde{\beta}$ as a function of $a/F$ and $F_c$. This is shown in Fig.~\ref{fig:tilt_angle_pcolormesh}. Again, we see that as $a/F$ increases, $\tilde{\beta}$ increases. Also, from the contours, we can see that $F_c^*$ decreases with increasing $a/F$. In Fig.~\ref{fig:tilt_angle_pcolormesh}, the top right corner is white. This white region corresponds to combinations of $a/F$ and $F_c$ where no physical solutions of $\tilde{\beta}$ exist. For large $F_c$, this happens when roughly $\tilde{\beta} \gtrsim \max{\tilde{\beta}}$, where $\max{\tilde{\beta}}$ is given by Eq.~(\ref{eq:max_beta_min}).

\begin{figure}[htbp]
\centering\includegraphics[width=0.65\textwidth]{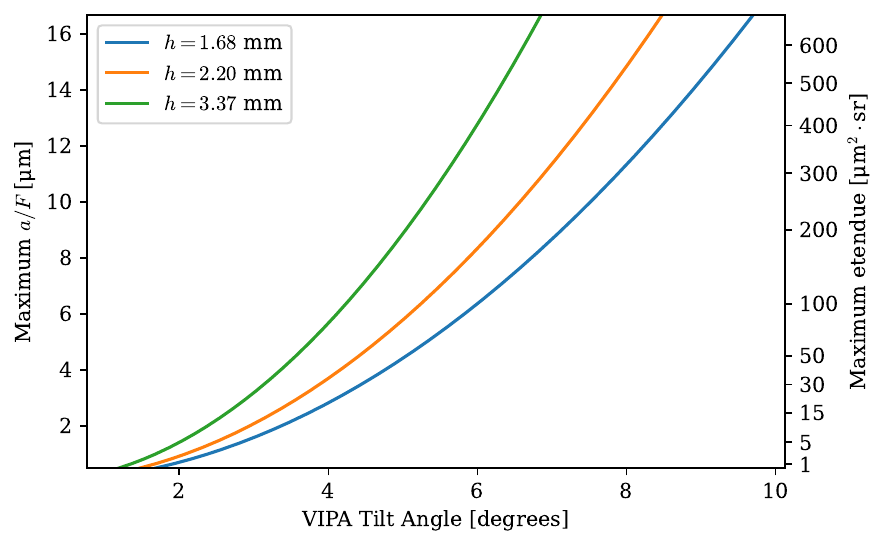}
\caption{Maximum $a/F$ and etendue $G_\mathrm{full}$ (Eq.~(\ref{eq:G_full}) with $F\gg1$ approximation) that can be accepted by a VIPA with no etendue coupling loss, given the VIPA tilt angle $\beta$. We show this for three different commercially-available values of VIPA thickness $h$ (from LightMachinery~Inc.). Here, we use $n'=1.4494$ (fused silica) and $n=1.0003$ (air). The quantity plotted on the horizontal axis is the $\tilde{\beta}(F_c^*)$ corresponding to the $a/F$ value on the vertical axis. }\label{fig:max_a_over_F}
\end{figure}

Another way to interpret the minimum VIPA tilt angle is to plot the maximum $a/F$ that can be accepted by a VIPA without etendue coupling loss, given a particular $\beta$, as shown in Fig.~\ref{fig:max_a_over_F}. In Fig.~\ref{fig:max_a_over_F}, the value on the horizontal axis is the $\tilde{\beta}(F_c^*)$ corresponding to the $a/F$ value on the vertical axis. We plot this for three different VIPA thicknesses $h$ that are commercially available. We can see that if the VIPA tilt angle $\beta$ is larger, then a larger $a/F$ can be accepted by the VIPA. Also, if the VIPA thickness $h$ is larger, then a larger $a/F$ can be accepted by the VIPA.

The quantity $a/F$ is a proxy for the etendue of the fiber, as we will now show. For a ray bundle propagating in a medium of refractive index $n$, the differential etendue associated with an area element $\mathrm{d}A$ and a solid angle element $\mathrm{d}\Omega$ is $\mathrm{d}G = n^2\,\mathrm{d}A\, \cos{\theta} \,\mathrm{d}\Omega$, where $\theta$ is the angle between the propagation direction and the normal to $\mathrm{d}A$ \cite{nonimagingoptics_chaves}. By integrating over the circular area of radius $a$, and the cone with half-angle $\arctan{(1/(2F))}$, we can compute the etendue of the fiber as:
\begin{equation}\label{eq:G_full}
    G_\mathrm{full} = \frac{n^2 \pi^2 a^2}{4F^2+1} \qquad\xrightarrow{F\gg1}\qquad  \approx\frac{n^2 \pi^2}{4}\left(\frac{a}{F}\right)^2
\end{equation}
Hence, $a/F$ is a proxy for the etendue of the fiber. Therefore, if the input optical fiber has a larger etendue, then the VIPA tilt angle $\beta$ must be larger if we do not want etendue coupling loss.

\section{Etendue coupled into VIPA}\label{sec:etendue_coupled}

In this section, we derive expressions for the 2D and 3D etendue that can be coupled into a VIPA, given $a$, $F$, $F_c$, $h$, $n$, $n'$, and $\beta$. \etalcite{Bouvet}{Bouvet2024} had derived an expression for the 3D etendue coupled into a VIPA, using the area-solid-angle geometrical formulation of etendue. Their analysis, however, made a small-angle approximation for the angular extent of the beam, effectively for $\theta_F$, and did not include the effect of refraction at the VIPA front surface. We instead derive the 2D and 3D coupled etendue from first principles using the phase-space formulation of etendue, and show explicit dependence on the experimentally-relevant physical parameters.

\subsection{Problem setup}\label{sec:etendue_problem_setup}

\subsubsection{Etendue definition and calculation approach}

\begin{figure}[htbp]
\centering\includegraphics[width=0.7\textwidth]{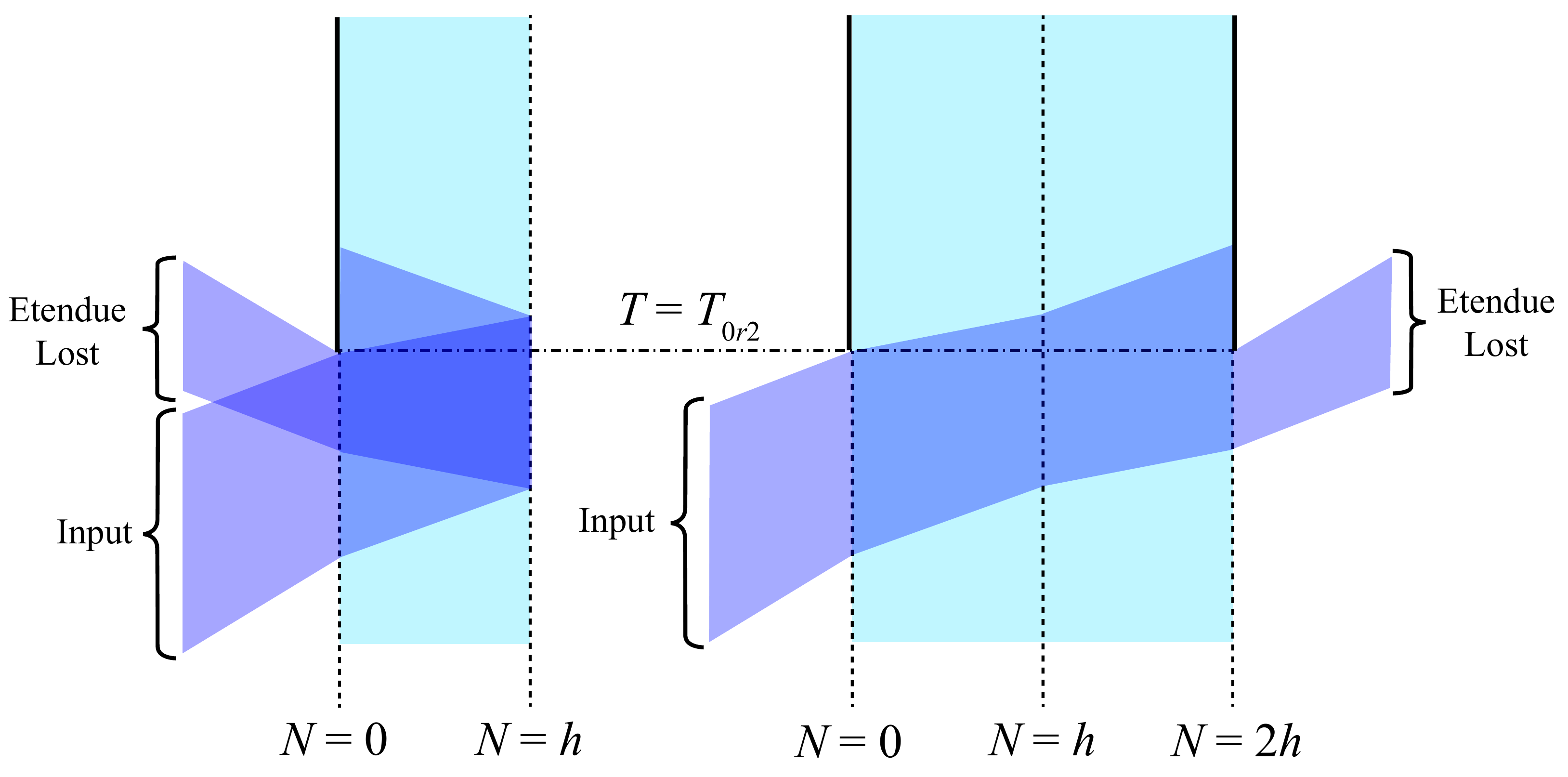}
\caption{Illustration of etendue lost when light is coupled into a VIPA whose tilt angle is less than $\tilde{\beta}$. At the VIPA HR surface ($N=2h$ in the right figure), rays with tangential coordinate less than $T_{0r2}$ can escape back through the entrance window. This light is lost.}\label{fig:VIPA_etendue_lost}
\end{figure}

In a 3D optical system, rays are described in a four-dimensional phase space $(x,y,p_x,p_y)$, where $x$ and $y$ are spatial coordinates, and $p_x=\tilde{n}\sin{\theta_x}$ and $p_y=\tilde{n}\sin{\theta_y}$ are the corresponding optical momenta, where $\tilde{n}$ is the refractive index of the medium. The differential etendue is \cite{nonimagingoptics_chaves,Winston2018}:
\begin{equation}\label{eq:G_def_3D_ps}
    \mathrm{d}G= \mathrm{d}x\,\mathrm{d}y\,\mathrm{d}p_x\,\mathrm{d}p_y
\end{equation}
For a 2D optical system, the corresponding reduced-phase-space differential etendue is \cite{nonimagingoptics_chaves}:
\begin{equation}\label{eq:G_def_2D_ps}
    \mathrm{d}G_\mathrm{2D}= \mathrm{d}x\,\mathrm{d}p_x
\end{equation}

Computing the etendue coupled into the VIPA involves carefully selecting the integrand and the bounds of integration over the phase space. Our approach is as follows. Within this section, we adopt a coordinate convention where the spatial coordinate $x$ in Eq.~(\ref{eq:G_def_3D_ps}) is in the $\hat{\mathbf{t}}$ direction, and the spatial coordinate $y$ in Eq.~(\ref{eq:G_def_3D_ps}) is in the $\hat{\mathbf{n}}\times\hat{\mathbf{t}}$ direction. We assume that there is a light source at the VIPA back surface, $N=h$, which is then propagated to the VIPA HR surface. Since a reflection occurs at $N=h$, the tangential coordinate of a certain ray at the VIPA HR surface is simply the tangential coordinate of the ray at $N=2h$. At $N=2h$, rays with a tangential coordinate below the bottom edge of the HR surface escape by traveling back through the VIPA entrance window. To compute the etendue coupled into the VIPA, we cut off these escaped rays in our calculation, which is represented mathematically by the integrand of the calculation. This geometry is illustrated in Fig.~\ref{fig:VIPA_etendue_lost}. The spatial and angular extent of the light source at $N=h$ are presented in Section~\ref{sec:G_extent}, with a full derivation in Supplement 1. In our calculations, we assume that the input fiber emits light uniformly in space, over a circle of radius $a$, and emits light uniformly in angle, over a cone with half-angle $\arctan{(1/(2F))}$.


\subsubsection{Spatial and angular extent of light source at $N=h$}\label{sec:G_extent}

The spatial extent of the light source at $N=h$ is set by the distance between the foci of the red and green ray bundles in Fig.~\ref{fig:VIPA_ray_expanded}. For the 3D etendue calculation, we take the spatial extent of the light source to be the ellipse $\mathcal{D}_A$:
\begin{subequations}
\begin{align}
    \mathcal{D}_A &\equiv \left\{ (x,y) \in \mathbb{R}^2 : \left(\frac{x-x_c}{R_x}\right)^2 + \left(\frac{y}{R_y}\right)^2 \leq 1 \right\} \label{eq:D_A_ellipse}\\
    x_c &= h \tan{\left(\arcsin{\left(\frac{n}{n'}\sin{\beta}\right)}\right)} \label{eq:x_c} \\
    R_x &= \frac{a}{\cos{\beta}} \sqrt{\frac{4F_c^2+1}{4F^2+1}} \label{eq:R_x} \\
    R_y &= a \label{eq:R_y}
\end{align}
\end{subequations}
When calculating the 2D etendue, we take the bounds of integration over $x$ to be $x_c-R_x$ and $x_c+R_x$. The angular extent of the light source at $N=h$ is set by the angular spread of the ray bundles in Fig.~\ref{fig:VIPA_ray_expanded}. For the 3D etendue calculation, we take the angular extent of the light source to be the optical momenta ellipse $\mathcal{D}_p$:
\begin{subequations}
\begin{align}
    \mathcal{D}_p &\equiv \left\{ (p_x,p_y) \in \mathbb{R}^2 : \left(\frac{p_x-p_{xc}}{p_{xR}}\right)^2 + \left(\frac{p_y}{p_{yR}}\right)^2 \leq 1 \right\} \label{eq:D_p_ellipse} \\
    p_{xc} &= n\sin{\beta}\cos{\theta_F} \label{eq:p_xc} \\
    p_{xR} &= n\cos{\beta}\sin{\theta_F} \label{eq:p_xR} \\
    p_{yR} &= n\sin{\left(\arctan{\left(\frac{1}{2F}\right)}\right)} \label{eq:p_yR}
\end{align}
\end{subequations}
When calculating the 2D etendue, we take the bounds of integration over $p_x$ to be $p_{xc}-p_{xR}$ and $p_{xc}+p_{xR}$. A full derivation of these quantities can be found in Supplement 1.

One point to note about the ellipses $\mathcal{D}_A$ and $\mathcal{D}_p$ is that these ellipses are defined hand-in-hand; the spatial and angular extents are connected. For example, since $p_{xR}$ in Eq.~(\ref{eq:p_xR}) uses the sine of the cone half-angle $\theta_F$ corresponding to $F_c$, therefore $R_x$ in Eq.~(\ref{eq:R_x}) is selected such that the Abbe sine condition\cite{Pedroti_Optics} is respected. In addition, $R_y$ in Eq.~(\ref{eq:R_y}) and $p_{yR}$ in Eq.~(\ref{eq:p_yR}) are selected in a way such that the etendue in the $y$-direction is conserved. For more details, see Supplement~1. 


\subsubsection{Ray 2 tangential coordinate $T_{0r2}$ at VIPA front surface}

Of particular importance is the tangential coordinate of ray 2 at the VIPA front surface, $T_{0r2}$. In order for the most etendue to be coupled into the VIPA, the VIPA and incoming beam should be oriented such that ray 2 clips the bottom edge of the HR surface at $N=0$, as illustrated in Fig.~\ref{fig:VIPA_ray_input}. In other words, $T_{0r2}$ is the tangential coordinate of the bottom edge of the HR surface. Hence, when the light source at $N=h$ propagates to the HR surface at $N=2h$, all rays with tangential coordinate less than $T_{0r2}$ will escape by traveling through the VIPA entrance window. Thus, $T_{0r2}$ will be used in the integrand of the etendue integral as a part of a cutoff condition.

Ray 2 has $\rho=D_c/2$ and $w=\frac{aF_c}{F}$. Using these quantities with Eq.~(\ref{eq:T_0r}), and the approximation $\frac{a}{f_\mathrm{coll}} \ll \frac{1}{2F_c}$, along with Eq.~(\ref{eq:d}) to eliminate $(f_c-d)$, we obtain:
\begin{equation}\label{eq:T0r2}
    T_{0r2} = h(\tan{\theta_{cr}'-\tan{\theta_{m1}'})} + F_c \left(\frac{a}{F}\right) \frac{\cos{\theta_F}}{\cos{(\beta-\theta_F)}}
\end{equation}
where $\theta_{m1}'$, $\theta_{cr}'$, and $\theta_F$ were given by Eqs.~(\ref{eq:thetap_m1}), (\ref{eq:thetap_cr}), and (\ref{eq:theta_F}) respectively. Notice that $T_{0r2}$ depends only on 5 parameters: $a/F$, $F_c$, $h$, $n/n'$, and $\beta$.


\subsection{2D etendue coupled into VIPA}\label{sec:2D_etendue_coupled}

Equiped with knowledge of the spatial and angular extent of the light source at $N=h$, we can now proceed to calculate the etendue coupled into the VIPA. Let us first consider the 2D etendue coupled into the VIPA. Let $x$ and $x'$ be the transverse coordinates at the planes $N=h$ and $N=2h$ respectively, where $x$ and $x'$ have the same direction as the VIPA tangential coordinate~$\hat{\mathbf{t}}$. Let the corresponding transverse optical momenta at $N=h$ and $N=2h$ be $p_x$ and $p_x'$ respectively. The mapping between the two planes is $(x,p_x)\mapsto(x',p_x')=(x+h\frac{p_x}{p_z(p_x)},\,p_x)$ where $p_z(p_x)=\sqrt{n'^2-p_x^2}$ is the longitudinal optical momentum. For light to stay inside the VIPA and not escape through the entrance window, we must have $x'\geq T_{0r2}$. This condition can be written as $x + \frac{hp_x}{\sqrt{n'^2-p_x^2}} - T_{0r2} \geq 0 $. In the etendue integral, this condition can be enforced by using the Heaviside step function, defined as:
\begin{equation}
    H(x) \equiv 
    \begin{cases} 
      1 & \mathrm{if}\:\:\: x\geq0 \\
      0 & \mathrm{if}\:\:\: x<0
   \end{cases}
\end{equation}
Then the 2D etendue coupled into the VIPA is:
\begin{subequations}\label{eq:2D_etendue_raw}
\begin{align}
    G_\mathrm{2D} &= \int_{p_-}^{p_+} \int_{x_-}^{x_{+}} H\left(x+\frac{hp_x}{\sqrt{n'^2-p_x^2}}-T_{0r2} \right) \, \mathrm{d}x \, \mathrm{d}p_x \label{eq:2D_etendue_raw_integral}
    \\
    &\begin{alignedat}{2}
        x_+ &\equiv x_c+R_x
        \qquad\qquad &
        x_- &\equiv x_c-R_x
        \\
        p_+ &\equiv p_{xc}+p_{xR}
        \qquad\qquad &
        p_- &\equiv p_{xc}-p_{xR}
    \end{alignedat}
\end{align}
\end{subequations}
where recall that $T_{0r2}$ is given by Eq.~(\ref{eq:T0r2}), and $x_c$, $R_x$, $p_{xc}$, and $p_{xR}$ are given by Eqs.~(\ref{eq:x_c}), (\ref{eq:R_x}), (\ref{eq:p_xc}), and (\ref{eq:p_xR}) respectively. Eq.~(\ref{eq:2D_etendue_raw}) has a closed-form solution, which is:
\begin{subequations}\label{eq:2D_etendue_solution}
\begin{align}
    G_\mathrm{2D} &=  (x_+ - T_{0r2}) (b_\mathrm{II} - a_\mathrm{II}) - h \left(\sqrt{n'^2-b_\mathrm{II}^2} - \sqrt{n'^2-a_\mathrm{II}^2}\right) + (x_+ - x_-) (b_\mathrm{III} - a_\mathrm{III}) \label{eq:2D_etendue_solution_alone}\\
    a_\mathrm{II} &\equiv \min{(\max{(p_-, p_{xt+})}, b_\mathrm{II})}
    \qquad\qquad\qquad 
    b_\mathrm{II} \equiv \min{(p_+, p_{xt-})} \label{eq:a_b_II} \\
    a_\mathrm{III} &\equiv \min{(\max{(p_-, p_{xt-})}, b_\mathrm{III})} 
    \qquad\qquad\qquad
    b_\mathrm{III} \equiv p_+ \label{eq:a_b_III} \\
    p_{xt+} &\equiv \frac{-n'(x_+ - T_{0r2})}{\sqrt{h^2 + (x_+ - T_{0r2})^2}}
    \qquad\qquad\qquad
    p_{xt-} \equiv \frac{-n'(x_- - T_{0r2})}{\sqrt{h^2 + (x_- - T_{0r2})^2}} \label{eq:p_xtpm}
\end{align}
\end{subequations}
A detailed derivation can be found in Supplement 1. Eq.~(\ref{eq:2D_etendue_solution}) is a closed-form expression for the 2D etendue coupled into a VIPA. If we trace back the definitions of the different constants in this expression, we can see that Eq.~(\ref{eq:2D_etendue_solution}) depends only on 7~parameters: $a$, $F$, $F_c$, $h$, $n$, $n'$, and $\beta$.


\subsection{3D etendue coupled into VIPA}\label{sec:3D_etendue_coupled}

Let us now consider the 3D etendue coupled into the VIPA. The approach is similar to the one we used for the 2D etendue in Section~\ref{sec:2D_etendue_coupled}, but here we consider two more dimensions for the phase space. Let $x$ and $y$ be transverse coordinates at the plane $N=h$, and $p_x$ and $p_y$ be the corresponding transverse optical momenta. Let $x'$ and $y'$ be transverse coordinates at the plane $N=2h$, and $p_x'$ and $p_y'$ be the corresponding transverse optical momenta. $x$ and $x'$ have the same direction as the VIPA tangential coordinate $\hat{\mathbf{t}}$, while $y$ and $y'$ have the same direction as $\hat{\mathbf{n}}\times\hat{\mathbf{t}}$. The mapping between the two planes is $(x,y,p_x,p_y)\mapsto(x',y',p_x',p_y')=(x+h\frac{p_x}{p_z(p_x,p_y)},\,y,\,p_x,\,p_y)$, where $p_z(p_x,p_y) = \sqrt{n'^2 - p_x^2 - p_y^2}$ is the longitudinal optical momentum. For light to stay inside the VIPA and not escape through the entrance window, we must have $x'\geq T_{0r2}$, which implies $x + \frac{h p_x}{p_z(p_x,p_y)} - T_{0r2} \geq 0$. Again, this condition can be enforced in the etendue integral by using the Heaviside step function as the integrand. Then the 3D etendue coupled into the VIPA is:
\begin{equation}\label{eq:3D_etendue_raw}
    G_\mathrm{3D}  = \iint_{\mathcal{D}_p} \iint_{\mathcal{D}_A} H\left(x + \frac{h p_x}{p_z(p_x,p_y)} - T_{0r2}\right) \, \mathrm{d}x \, \mathrm{d}y \, \mathrm{d}p_x \, \mathrm{d}p_y
\end{equation}
where $\mathcal{D}_A$ and $\mathcal{D}_p$ are ellipses defined by Eqs.~(\ref{eq:D_A_ellipse}) and (\ref{eq:D_p_ellipse}) respectively. Eq.~(\ref{eq:3D_etendue_raw}) is a quadruple integral that can be reduced to a double integral over polar coordinates. The reduced version is:
\begin{subequations}\label{eq:3D_etendue_solution}
\begin{align}
    G_\mathrm{3D} &= R_x R_y p_{xR} p_{yR} \int_0^{2\pi} \int_0^1 \Phi_\mathrm{R}\left(\frac{T_{0r2} - hp_s(r,\varphi) -x_c}{R_x}\right)\, r\, \mathrm{d}r\,\mathrm{d}\varphi \label{eq:3D_etendue_solution_alone}\\
    \Phi_\mathrm{R}(z) &\equiv\begin{cases} 
      \pi & \mathrm{if}\:\:\: z\leq-1 \\
      \arccos{z}-z\sqrt{1-z^2} & \mathrm{if}\:\:\: -1<z<1 \\
      0 & \mathrm{if}\:\:\: z\geq1
   \end{cases}\label{eq:Phi_R}\\
   p_s(r,\varphi) &\equiv \frac{(p_{xc}+p_{xR}r\cos{\varphi})}{\sqrt{n'^2-(p_{xc}+p_{xR}r\cos{\varphi})^2-(p_{yR}r\sin{\varphi})^2}} \label{eq:p_s}
\end{align}
\end{subequations}
A detailed derivation can be found in Supplement 1. Eq.~(\ref{eq:3D_etendue_solution}) is an expression for the 3D etendue coupled into a VIPA. Eq.~(\ref{eq:3D_etendue_solution}) can be easily and quickly evaluated numerically, as it is an integral over the unit disk. In Eq.~(\ref{eq:3D_etendue_solution}), recall that the constants $T_{0r2}$, $x_c$, $R_x$, $R_y$, $p_{xc}$, $p_{xR}$, and $p_{yR}$ are given by Eqs.~(\ref{eq:T0r2}), (\ref{eq:x_c}), (\ref{eq:R_x}), (\ref{eq:R_y}) (\ref{eq:p_xc}), (\ref{eq:p_xR}), and (\ref{eq:p_yR}) respectively. If we trace back the definitions of these different constants in the expression, we can see that Eq.~(\ref{eq:3D_etendue_solution}) depends only on 7~parameters: $a$, $F$, $F_c$, $h$, $n$, $n'$, and $\beta$.


\subsection{Etendue coupling efficiency}

\begin{figure}[htbp]
\centering\includegraphics[width=0.55\textwidth]{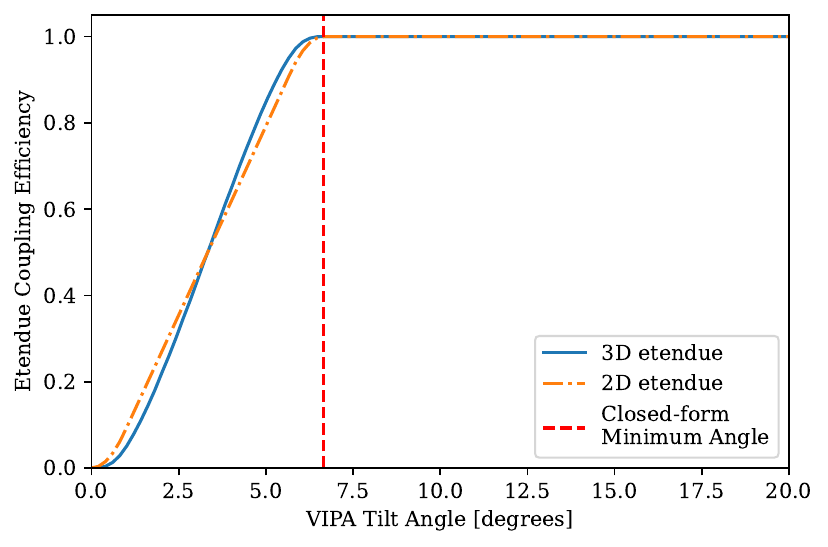}
\caption{2D and 3D etendue coupling efficiency of a VIPA as a function of VIPA tilt angle $\beta$. Here, we use $a=25$ \textmu m, $F=5$, $F_c=5$, $h=2.2$ mm, $n'=1.4494$ (fused silica), and $n=1.0003$ (air). The solid blue curve is $\eta$, computed from Eq.~(\ref{eq:eta}), and the dotted-dashed orange curve is $\eta_\mathrm{2D}$, computed from Eq.~(\ref{eq:eta2D}). The vertical dashed red line is the minimum VIPA tilt angle for no etendue coupling loss, $\tilde{\beta}$, computed from Eq.~(\ref{eq:beta_min}). Notice that the etendue coupling efficiency is $1$ when $\beta\geq\tilde{\beta}$.}\label{fig:etendue_eta}
\end{figure}

With expressions for the 2D and 3D coupled etendue, we can then obtain the etendue coupling efficiency. This is the ratio of the etendue that stays in the VIPA to the etendue that enters the VIPA, ignoring the reflectances of the internally reflective surfaces. Let $\eta$ and $\eta_\mathrm{2D}$ respectively be the 3D and 2D etendue coupling efficiencies of a VIPA in a particular configuration. These are:
\begin{subequations}
\begin{align}
    \eta &= \frac{G_\mathrm{3D}}{G_\mathrm{full}} \label{eq:eta}\\
    \eta_\mathrm{2D} &= \frac{G_\mathrm{2D}}{G_\mathrm{full,2D}} \label{eq:eta2D}
\end{align}
\end{subequations}
where $G_\mathrm{3D}$ is given by Eq.~(\ref{eq:3D_etendue_solution}), $G_\mathrm{full}$ is given by Eq.~(\ref{eq:G_full}), $G_\mathrm{2D}$ is given by Eq.~(\ref{eq:2D_etendue_solution}), and $G_\mathrm{full,2D}$ is the 2D etendue of the optical fiber, which is:
\begin{equation}
    G_\mathrm{full,2D} = \frac{4na}{\sqrt{4F^2+1}}
\end{equation}

Fig.~\ref{fig:etendue_eta} plots the etendue coupling efficiency as a function of VIPA tilt angle $\beta$, for a particular configuration. When $\beta=0$, the etendue coupling efficiency is $0$. Notice that when $\beta$ is larger than the minimum VIPA tilt angle for no etendue coupling loss, $\tilde{\beta}$ (Eq.~(\ref{eq:beta_min})), then the etendue coupling efficiency is $1$. Hence, our calculations the for the coupled etendue are consistent with the derived $\tilde{\beta}$ from Section~\ref{sec:tilt_angle}.


\subsection{Effect on optical fiber modal information and modal noise}

The presence of an etendue coupling efficiency in the VIPA suggests that some of the fiber's modal information can be affected by the VIPA when $\beta<\tilde{\beta}$. This is because the etendue is related to the number of modes of an optical system, in that the number of transverse modes occupied by a beam with etendue $G$ and wavelength~$\lambda$ is $G/\lambda^2$ \cite{Brooker_Optics}. However, one should be careful to note that the VIPA's etendue coupling efficiency $\eta$ should not be interpreted as the fraction of fiber modes transmitted through the VIPA. A reduction in the transmitted etendue does not in general correspond to the blockage of a proportional number of fiber modes.

To see why, let us consider how light propagates through the optical system and the role of the VIPA within this system. Light exits the input fiber and is directed by the collimator and cylindrical lens into the VIPA. When $\eta<1$, the VIPA effectively acts as a partial obscuration at $N=2h$, blocking the part of the etendue phase space $\mathcal{D}_A\cup\mathcal{D}_p$ where $x'<T_{0r2}$. This optical system, comprising the collimator, the cylindrical lens, and the VIPA obscuration, can be considered a linear optical device. Now, according to the mode-converter formalism by Miller~\cite{Miller2012,Miller2019}, any linear optical device can be represented as an operator $\mathsf{D}$, whose singular value decomposition identifies a set of orthogonal input modes that are coupled to a set of orthogonal output modes, with the singular values quantifying the coupling strengths between pairs of modes in each set. These modes are determined by the optical device and do not need to be the same as the fiber modes. Hence, the VIPA obscuration does not simply select and block a subset of fiber modes. Instead, the VIPA obscuration changes $\mathsf{D}$, and therefore changes the way the input field is coupled through the optical device. Hence, the fraction of etendue clipped by the VIPA obscuration is not the same as the fraction of fiber modes blocked. However, this does not mean that the fiber's modal information is unaffected by the VIPA obscuration. An obscuration may weaken the coupling strengths between the input and output modes to an extent such that there is a reduction in the the number of usable output modes accessible above a specified noise threshold \cite{Piestun2000}. Hence, while the VIPA etendue coupling efficiency $\eta$ is not the same as the fraction of fiber modes transmitted through the VIPA, a reduction in $\eta$ can reduce the amount of fiber modal information accessible at a certain signal-to-noise ratio. More investigation is required in order to determine the effect of VIPA etendue coupling loss on fiber modal information.

One problem with the use of multimode optical fibers is the phenomenon of fiber modal noise~\cite{Epworth1979}, which is of particular relevance to multimode fiber-fed astronomical spectroscopy~\cite{Baudrand2001,Lemke2011,Leung2025MS}. Modal noise is a measurement noise that arises from the high-contrast speckle pattern seen in the near-field and far-field of multimode fibers, which is caused by the interference of the fiber modes. According to \etalcite{Lemke}{Lemke2011}, a truncation in the beam of a spectrometer, downstream of the fiber, can reduce the measurement signal-to-noise ratio of the spectrometer. Since the VIPA is downstream of the fiber and acts as a partial obscuration when $\eta<1$, it can therefore reduce the instrument signal-to-noise ratio. Hence, based on this interpretation, in order to reduce the impacts of modal noise, the VIPA should be operated in a configuration where $\eta=1$. However, more investigation, particularly experimental validation, is required in order to ascertain this.


\section{VIPA transmission}\label{sec:transmission}

In this section, we derive the transmission of a VIPA, given $a$, $F$, $F_c$, $h$, $n$, $n'$, $\beta$, and the VIPA length $L$. We define the ``transmission'' as the ratio of the intensity of light exiting the VIPA through the PR surface to the intensity of light entering the VIPA through the entrance window. The ``transmission'' that we derive here refers to the geometrical incoherent transmission, rather than the wavelength-dependent coherent transmission at a specific VIPA output angle.

\subsection{Transmission computed from reflectances alone}

We first consider calculations of the VIPA transmission using the reflectances. Recall that we denoted $R_1$ and $R_2$ as the internal reflectances of the front and back surfaces of the VIPA respectively. We ignore the entrance window reflectance. If we assume that the VIPA is infinitely long, then we can calculate the VIPA transmission by considering the attenuation of intensity by the reflectances $R_1$ and $R_2$ in each round trip. Then the VIPA transmission can be found by evaluating the following geometric series:
\begin{equation}\label{eq:T_VIPA_gs}
    T_\mathrm{VIPA} = \sum_{l=0}^\infty (1-R_2)(R_1R_2)^l = \frac{1-R_2}{1-R_1R_2}
\end{equation}
where $(1-R_2)$ is the transmittance of the PR surface. \etalcite{Weiner}{Weiner2012} also arrived at this same result for the VIPA transmission, instead by taking the transmission averaged over a phase shift. However, in reality, the VIPA is not infinitely long. Its finite length leads to important effects, such as a decrease in resolving power \cite{Hu2015}. \etalcite{Hu}{Hu2015} considered this in their VIPA spectral dispersion law, noting that for a VIPA of length $L$, the number of round-trip reflections inside the VIPA is roughly:
\begin{equation}
    M \approx \frac{L}{2h\tan{\beta'}}
\end{equation}
where $\beta'=\arcsin{\left(\frac{n}{n'}\sin{\beta} \right)}$. If light makes $M$ bounces off of the VIPA front surface, then we can modify Eq.~(\ref{eq:T_VIPA_gs}) to write:
\begin{equation}\label{eq:T_VIPA_gs_trunc}
    T_\mathrm{VIPA} = \sum_{l=0}^M (1-R_2)(R_1R_2)^l = \frac{(1-R_2)(1-(R_1R_2)^{M+1})}{1-R_1R_2}
\end{equation}

While this description of transmission in Eq.~(\ref{eq:T_VIPA_gs_trunc}) is more accurate than Eq.~(\ref{eq:T_VIPA_gs}), it is still missing some effects. Namely, it neglects etendue coupling losses back through the entrance window, which we showed are important in Section \ref{sec:etendue_coupled}. It also neglects etendue losses at the top of the VIPA. By the time the beam propagates to the top of the VIPA, it will be partially clipped by the VIPA top edge, only allowing certain parts of the beam to be transmitted. We will derive this transmitted etendue in Section \ref{sec:T_etendue_round_trip}.


\subsection{Transmitted etendue after a certain number of round trips inside the VIPA}\label{sec:T_etendue_round_trip}

Let $L$ specifically be the length of the HR-coated surface of the VIPA. Then the VIPA length, as measured from tangential coordinate $T=0$, is:
\begin{equation}\label{eq:tilde_L}
    \tilde{L}\equiv L+T_{0r2}
\end{equation}
Consider the part of the beam that makes $l\in\mathbb{Z}_{>0}$ round trips inside the VIPA after the first reflection off of the PR surface. After $l$ round trips, this part of the beam reaches the VIPA PR surface, having traveled a distance of $\Delta N=2hl$ along direction $\hat{\mathbf{n}}$. The part of the beam that is transmitted through the PR surface, and not clipped by the top of the VIPA, will satisfy $x+\frac{(\Delta N)p_x}{p_z(p_x,p_y)} \leq \tilde{L}$, in addition to satisfying the etendue coupling condition, $x + \frac{h p_x}{p_z(p_x,p_y)} - T_{0r2} \geq 0$, from earlier in Section \ref{sec:3D_etendue_coupled}. To calculate the etendue transmitted, we use a similar approach as in Section \ref{sec:3D_etendue_coupled}, but we modify the integrand in Eq.~(\ref{eq:3D_etendue_raw}) so that it incorporates both inequalities. Then the part of the beam that travels $\Delta N$, after the first reflection off the PR surface, has the following transmitted 3D etendue through the VIPA, which we denote as $G_\mathrm{RT}$:
\begin{equation}\label{eq:G_RT_raw}
    G_\mathrm{RT} = \iint_{\mathcal{D}_p} \iint_{\mathcal{D}_A} H\left(x + \frac{h p_x}{p_z(p_x,p_y)} - T_{0r2}\right) \, H\left(\tilde{L} - \left[x+\frac{(\Delta N) p_x}{p_z(p_x,p_y)} \right]\right) \, \mathrm{d}x \, \mathrm{d}y \, \mathrm{d}p_x \, \mathrm{d}p_y
\end{equation}
Like Eq.~(\ref{eq:3D_etendue_raw}), Eq.~(\ref{eq:G_RT_raw}) is a quadruple integral that can be reduced to a double integral over polar coordinates. The reduced version is:
\begin{subequations}\label{eq:G_RT_solution}
\begin{align}
    G_\mathrm{RT} &= R_x R_y p_{xR} p_{yR} \int_0^{2\pi} \int_0^1 I_\mathrm{RT}(r,\varphi)\, r\, \mathrm{d}r\,\mathrm{d}\varphi \label{eq:G_RT_solution_alone}\\
    I_\mathrm{RT}(r,\varphi) &\equiv 
    \max{\left(\Phi_\mathrm{L}\left(\frac{\tilde{L}-(\Delta N)p_s(r,\varphi)-x_c}{R_x}\right)
    -\Phi_\mathrm{L}\left(\frac{T_{0r2}-hp_s(r,\varphi)-x_c}{R_x}\right),0\right)} \label{eq:I_RT}\\
    \Phi_\mathrm{L}(z) &\equiv\begin{cases} 
      0 & \mathrm{if}\:\:\: z\leq-1 \\
      \pi-\arccos{z}+z\sqrt{1-z^2} & \mathrm{if}\:\:\: -1<z<1 \\
      \pi & \mathrm{if}\:\:\: z\geq1 \label{eq:Phi_L}
   \end{cases}
\end{align}
\end{subequations}
where recall that $p_s(r,\varphi)$ is Eq.~(\ref{eq:p_s}). A detailed derivation can be found in Supplement 1. Eq.~(\ref{eq:G_RT_solution}) can be easily and quickly evaluated numerically, as it is an integral over the unit disk.


\subsection{Total transmitted etendue and VIPA transmission}

Eq.~(\ref{eq:G_RT_solution}) describes the etendue transmitted by the part of the beam that makes $l$ round trips inside the VIPA after the first reflection off of the PR surface. However, if $l$ is sufficiently small, then the beam is guaranteed to satisfy the clipping condition $x+\frac{(\Delta N)p_x}{p_z(p_x,p_y)} \leq \tilde{L}$, and we do not need to worry about clipping at the top of the VIPA. We denote this threshold for $l$ as $\tilde{M}_r$, which is:
\begin{equation}\label{eq:M_tilde_r}
    \tilde{M}_r = \left\lfloor\frac{\tilde{L}-\left(h \tan{\left(\arcsin{\left(\frac{n}{n'}\sin{\beta}\right)}\right)} + \left(\frac{a}{F}\right) \frac{F_c}{\cos{\beta}} \right)}{2h\tan{\left(\arcsin{\left(\frac{n}{n'}\sin{(\beta+\theta_F)}\right)} \right)}}\right\rfloor
\end{equation}
A full derivation can be found in Supplement 1. $\tilde{M}_r$ is defined such that for $l\leq \tilde{M}_r$, the corresponding part of the beam is not clipped by the top of the VIPA.

\begin{figure}[htbp]
\centering\includegraphics[width=0.7\textwidth]{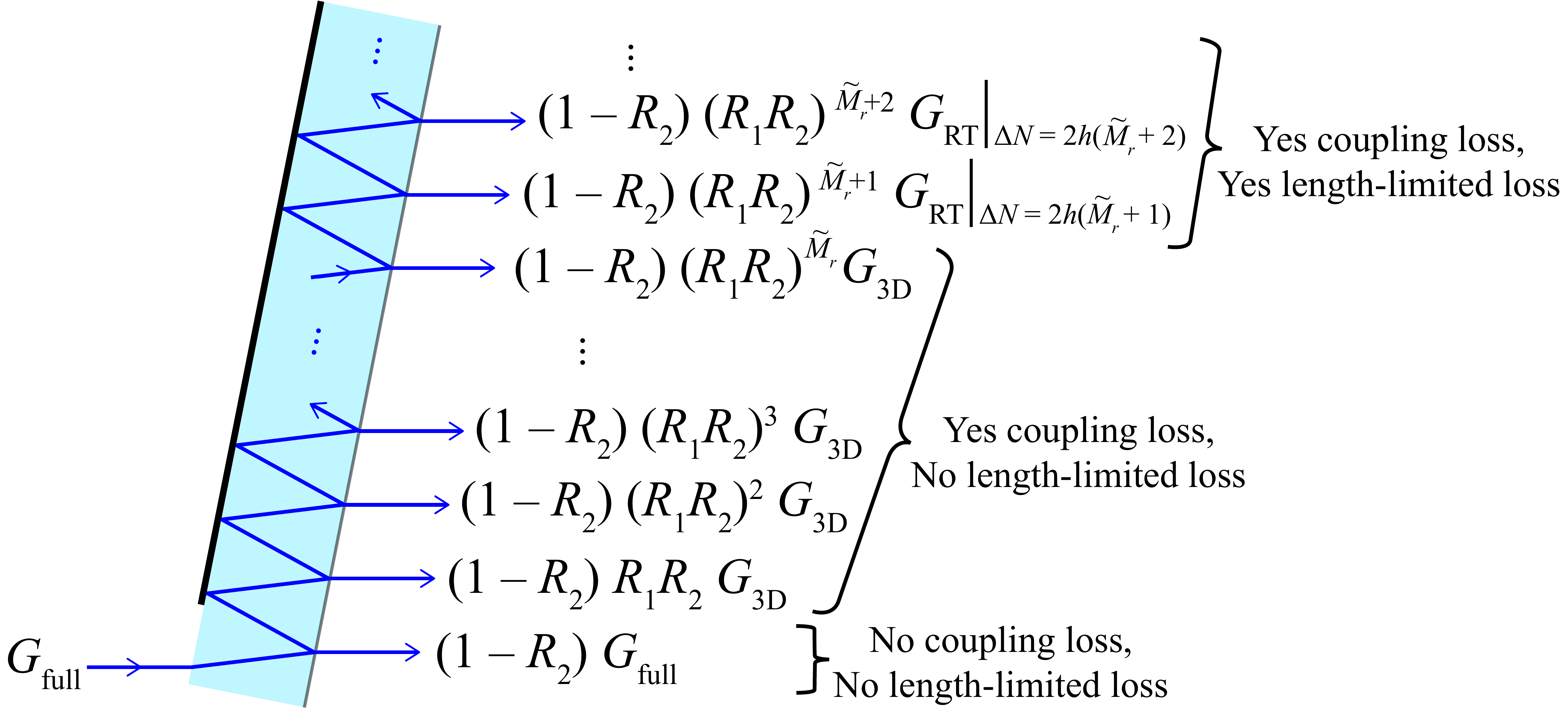}
\caption{Illustration of the three contributions to the transmitted etendue of a VIPA.}\label{fig:G_out}
\end{figure}

We can now compute the total transmitted etendue through the VIPA and hence the VIPA transmission. Consider Fig.~\ref{fig:G_out}. There are three contributions to the transmitted etendue. First, when the beam reaches the PR surface for the very first time, $(1-R_2)G_\mathrm{full}$ is transmitted out of the VIPA, experiencing no etendue coupling loss. $G_\mathrm{full}$ is Eq.~(\ref{eq:G_full}). Second, for subsequent round trips where $l\leq \tilde{M}_r$, the beam will have experienced etendue coupling loss, but not length-limited clipping loss. The correpsonding transmitted etendue contribution is $(1-R_2)(R_1R_2)^l G_\mathrm{3D}$, where $G_\mathrm{3D}$ is Eq.~(\ref{eq:3D_etendue_solution}). Third, for subsequent round trips where $l> \tilde{M}_r$, the beam will have experienced etendue coupling loss and length-limited clipping loss. The corresponding transmitted etendue contribution is $(1-R_2)(R_1R_2)^l G_\mathrm{RT}\bigr\rvert_{\Delta N=2hl}$, where $G_\mathrm{RT}$ is Eq.~(\ref{eq:G_RT_solution}). Each $l$ can be thought of as corresponding to one virtual source. Summing all of these contributions, we find that the transmitted etendue is:
\begin{equation}\label{eq:G_out}
    G_\mathrm{out} = (1-R_2)\left(G_\mathrm{full}+G_\mathrm{3D} \frac{(R_1R_2)^{\tilde{M}_r+1}-R_1R_2}{R_1R_2-1}
    + \sum_{l=\tilde{M}_r+1}^\infty (R_1R_2)^l G_\mathrm{RT}\biggr\rvert_{\Delta N=2hl}
    \right)
\end{equation}
Mathematically, we are allowed to have a summation to infinity in Eq.~(\ref{eq:G_out}), because $G_\mathrm{RT}$ will evaluate to 0 anyway when $l$ is sufficiently large. However, when evaluating Eq.~(\ref{eq:G_out}) numerically, we limit the summation to a more reasonable finite number, just several times that of $\tilde{M}_r$.

With Eq.~(\ref{eq:G_out}), we can then find the VIPA transmission $T_\mathrm{VIPA}$, which is:
\begin{equation}\label{eq:T_VIPA}
    T_\mathrm{VIPA} = \frac{G_\mathrm{out}}{G_\mathrm{full}}
\end{equation}
If we trace back the definitions of the different constants, we can see that the transmitted etendue Eq.~(\ref{eq:G_out}) and hence the VIPA transmission Eq.~(\ref{eq:T_VIPA}) depend on 8~parameters: $a$, $F$, $F_c$, $h$, $n$, $n'$, $\beta$, and $L$.

\section{Validation with Zemax OpticStudio}\label{sec:Zemax_sim}

In this section, we validate our mathematical models derived in Sections \ref{sec:tilt_angle}, \ref{sec:etendue_coupled}, and \ref{sec:transmission} using ray-tracing simulations in Ansys Zemax OpticStudio (version 2025 R2.04) non-sequential mode. We ran simulations for 9 different combinations of fiber radius $a$ and cylindrical lens focal ratio~$F_c$. 

\subsection{Optical fiber and VIPA coupling optics}\label{sec:Zemax_sim_fiber_and_coupling}

We modeled the input optical fiber using the ``Source Two Angle'' object in Zemax non-sequential mode. We set the source's X and Y half-widths to our desired $a$, and the source's X and Y half-angles to $\arctan{(1/(2F))}$. Since we assumed a circular-core optical fiber, we set both the spatial and angular shape of the source to be elliptical. We also set the source's irradiance to be uniform in angle space.

For optical system that couples light into the VIPA, we used Zemax models of commercial off-the-shelf lenses from Thorlabs. It should be noted that the VIPA interference pattern is destroyed if we use paraxial lens objects in Zemax, and so we used actual lens models. The lens models we used for the collimator and cylindrical lens in our simulations are detailed in Supplement 1. In our simulations, we used $F_c$ values of 5, 10, and 12.5, and a wavelength of $\lambda=1083$ nm. We kept $F=5$ for the fiber source, so that the diameter of the collimated beam after the collimator is roughly $D_c\approx20$~mm. We used $a$ values of $25$~\textmu m, $50$~\textmu m, and $100$~\textmu m.


\subsection{Zemax VIPA model}\label{sec:Zemax_sim_VIPA}

\begin{figure}[htbp]
\centering\includegraphics[width=0.37\textwidth]{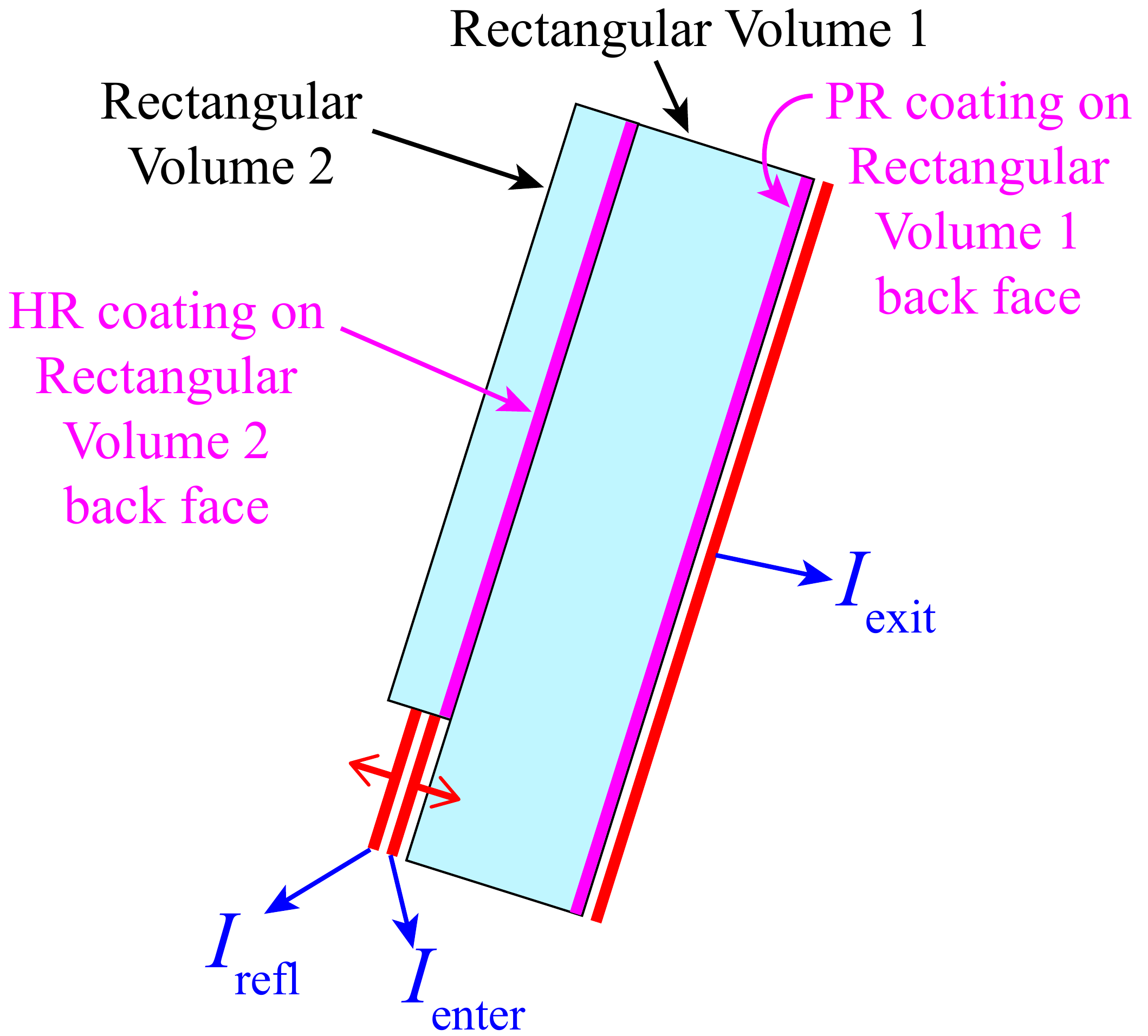}
\caption{Illustration of our VIPA model in Zemax non-sequential mode. The VIPA model is comprised of two rectangular volumes. The back faces of both rectangular volume 1 and rectangular volume 2 are coated with the PR and HR coating respectively. We placed three detectors around the VIPA to measure coupling efficiency and transmission. The thickness of rectangular volume~2 is exaggerated here for illustrative purposes.}\label{fig:Zemax_VIPA_geometry}
\end{figure}

We modeled the VIPA using two rectangular volume objects in Zemax. This is illustrated in Fig.~\ref{fig:Zemax_VIPA_geometry}. Rectangular volume~1 is the VIPA medium itself. In our simulations, we set the material to be fused silica ($n'=1.4494$ at $\lambda=1083$ nm), and the thickness to be $h=2.20$~mm. We coat the back face of rectangular volume~1 with the PR coating, which in our simulations we use $R_2=95\%$. We coat the front face of rectangular volume~1 with a 100\% transmissive coating. Rectangular volume~2 is a placeholder for the HR coating. The length of rectangular volume~2 is $L$, which recall from Section~\ref{sec:T_etendue_round_trip} is the length of the HR-coated surface of the VIPA. We take $L=21$~mm in our simulations. We coat the back face of rectangular volume~2 with the HR surface, which in our simulations we use $R_1=99.5\%$. Notice in Fig.~\ref{fig:Zemax_VIPA_geometry} that rectangular volume~2 is shorter than rectangular volume~1. This is to allow for the transmissive entrance window. This VIPA modeling approach with two rectangular volumes was also used by \etalcite{Aryana}{Aryana2026}, which has been experimentally validated.

We placed three ``Detector Rectangle'' objects around the VIPA to measure the incoherent irradiance. These three detectors are illustrated in red in Fig.~\ref{fig:Zemax_VIPA_geometry}. The detectors capture the irradiance entering the VIPA ($I_\mathrm{enter}$), the irradiance exiting the VIPA ($I_\mathrm{exit}$), and the irradiance that escapes back through the VIPA entrance window because of etendue coupling loss ($I_\mathrm{refl}$). The two detectors for $I_\mathrm{enter}$ and $I_\mathrm{refl}$ are directional, only counting rays that strike the detector from one side, as indicated by the red arrows in Fig.~\ref{fig:Zemax_VIPA_geometry}. Using these detectors, we can measure the etendue coupling efficiency to be:
\begin{equation}\label{eq:eta_zemax}
    \eta_\mathrm{Zemax} = \frac{I_\mathrm{enter}-I_\mathrm{refl}/R_2}{I_\mathrm{enter}}
\end{equation}
and we can measure the VIPA transmission to be:
\begin{equation}\label{eq:T_VIPA_Zemax}
    T_{\mathrm{VIPA,Zemax}} = \frac{I_\mathrm{exit}}{I_\mathrm{enter}}
\end{equation}


\subsection{Zemax simulation}\label{sec:Zemax_sim_actual}

In the Zemax system explorer's non-sequential mode settings, we set the ``maximum intersections per ray'' to be 4000 and the ``minimum relative ray intensity'' to be $10^{-8}$. The former and latter must be large enough and small enough respectively, so that rays can reach the top of the VIPA. In each of our simulations, we used $10^5$ analysis rays.

\begin{figure}[htbp]
  \centering
  \begin{subfigure}{\textwidth}
    \centering
    \includegraphics[width=\textwidth]{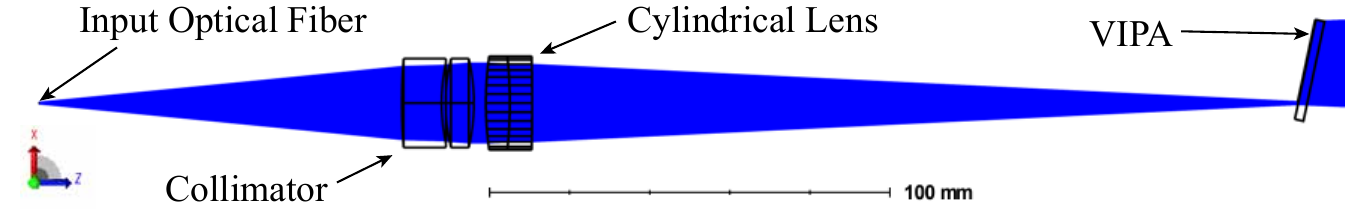}
    \caption{Zemax 3D layout.}
    \label{fig:F10_NSC3DLayout1_rasterized}
  \end{subfigure}
  \hfill
  \begin{subfigure}{0.39\textwidth}
    \centering
    \includegraphics[width=\textwidth]{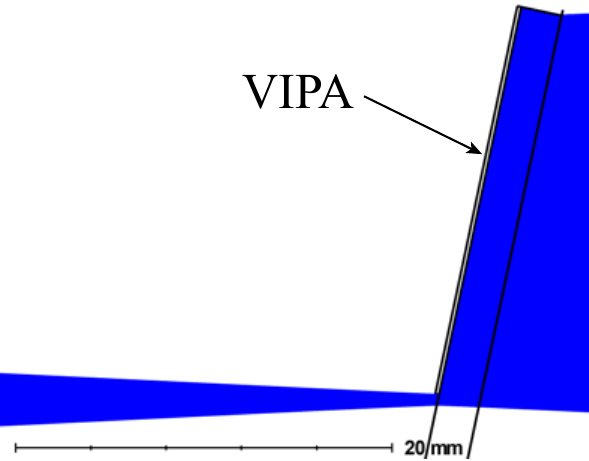}
    \caption{Close-up of the VIPA.}
    \label{fig:F10_NSC3DLayout2_rasterized}
  \end{subfigure}
  \hfill
  \begin{subfigure}{0.55\textwidth}
    \centering
    \includegraphics[width=0.95\textwidth]{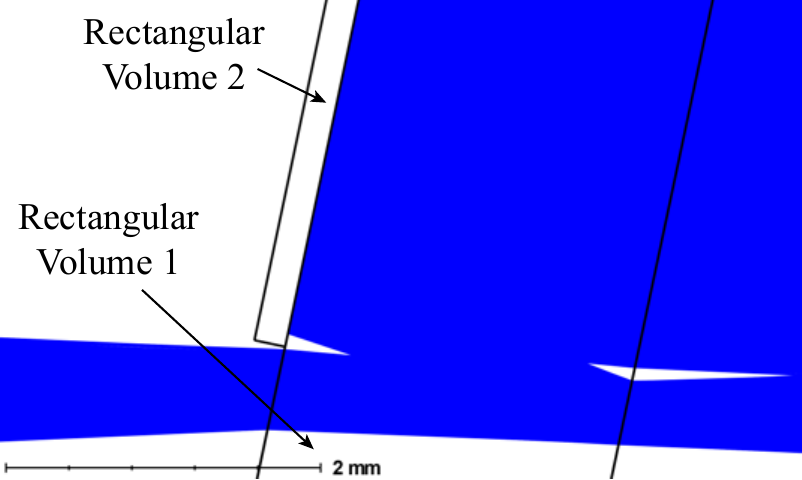}
    \caption{Close-up showing the VIPA coupling and rectangular volumes.}
    \label{fig:F10_NSC3DLayout3_rasterized}
  \end{subfigure}
  \caption{Zemax non-sequential mode 3D layout of one simulation. Here, the collimator is the Thorlabs ACA254-100-B air-spaced doublet and the cylindrical lens is the Thorlabs ACY254-200-B cylindrical achromat. Our simulation parameters here are $\beta=12^\circ$, $a=100$~\textmu m, $F=5$, $F_c=10$, $h=2.20$~mm, $n'=1.4494$ (fused silica), $n=1.0003$ (air), $R_1=99.5\%$, $R_2=95\%$, and $L=21$~mm.}
  \label{fig:F10_NSC3DLayout}
\end{figure}

For each $(a,F_c)$ pair, we swept $\beta$ from $0^\circ$ to $12^\circ$, and ran a non-sequential ray trace simulation at each $(a,F_c,\beta)$ triplet. We made sure that non-sequential rays were split. In each of these simulations, we positioned the VIPA such that the incoming beam forms focus at the VIPA back surface, and such that the incoming beam enters the VIPA right below where the HR surface ends. The former and latter were done by first calculating the appropriate values of $d$ and $T_{0r2}$ respectively, given by Eqs.~(\ref{eq:d}) and (\ref{eq:T0r2}) respectively, and then offsetting the VIPA accordingly. Fig.~\ref{fig:F10_NSC3DLayout} shows an example of one such simulation. We used the Python ZOSPy library\cite{vanVught2024_ZOSPy}, a wrapper of the Zemax ZOS-API, to efficiently sweep over $\beta$ values and position the VIPA accordingly.

\begin{figure}[htbp]
  \centering
  \begin{subfigure}{\textwidth}
    \centering
    \includegraphics[width=\textwidth]{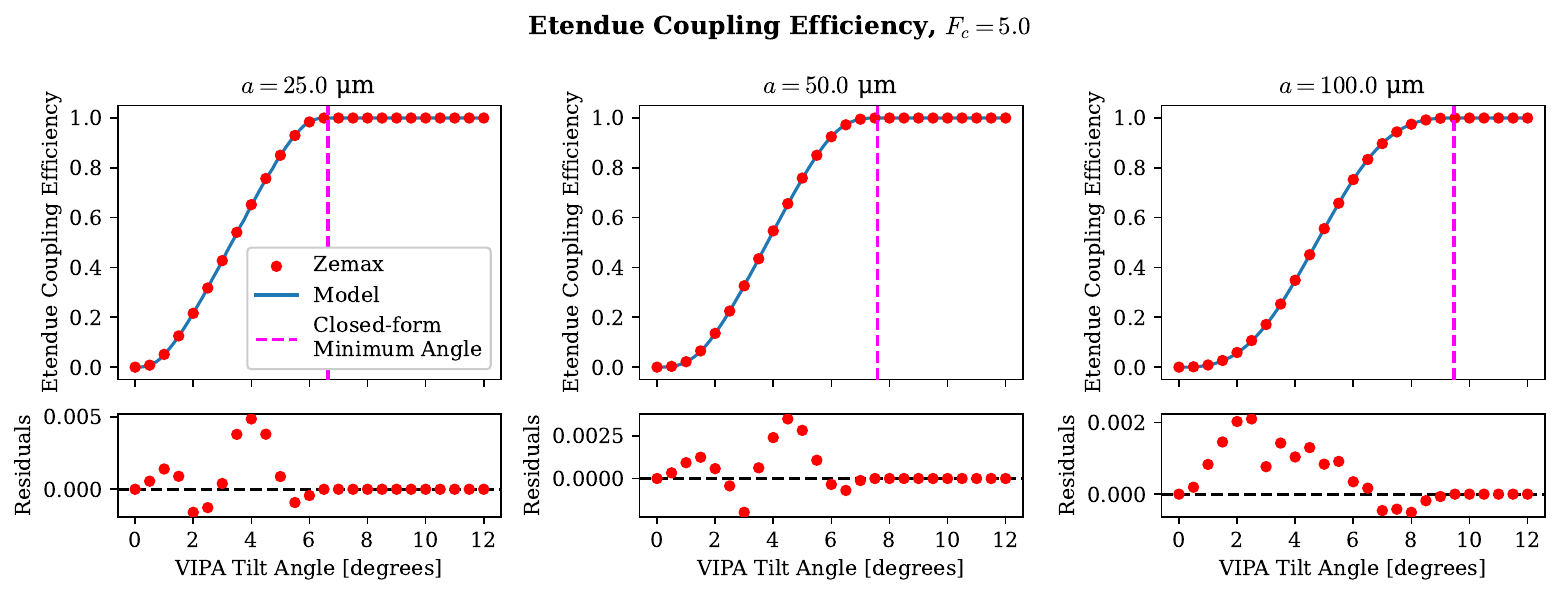}
    \caption{$\eta$ for $F_c=5$}
    \label{fig:etendue_eta_Fc5.0}
  \end{subfigure}
  \hfill
  \begin{subfigure}{\textwidth}
    \centering
    \includegraphics[width=\textwidth]{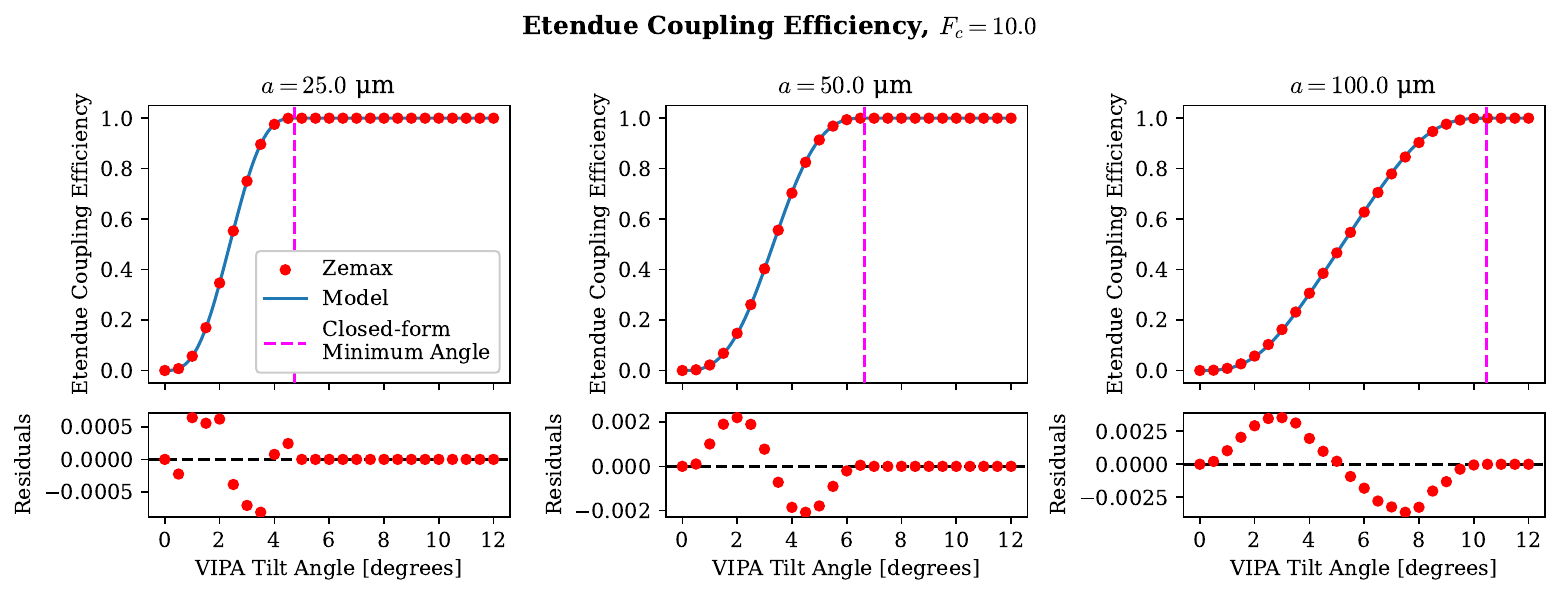}
    \caption{$\eta$ for $F_c=10$}
    \label{fig:etendue_eta_Fc10.0}
  \end{subfigure}
  \hfill
  \begin{subfigure}{\textwidth}
    \centering
    \includegraphics[width=\textwidth]{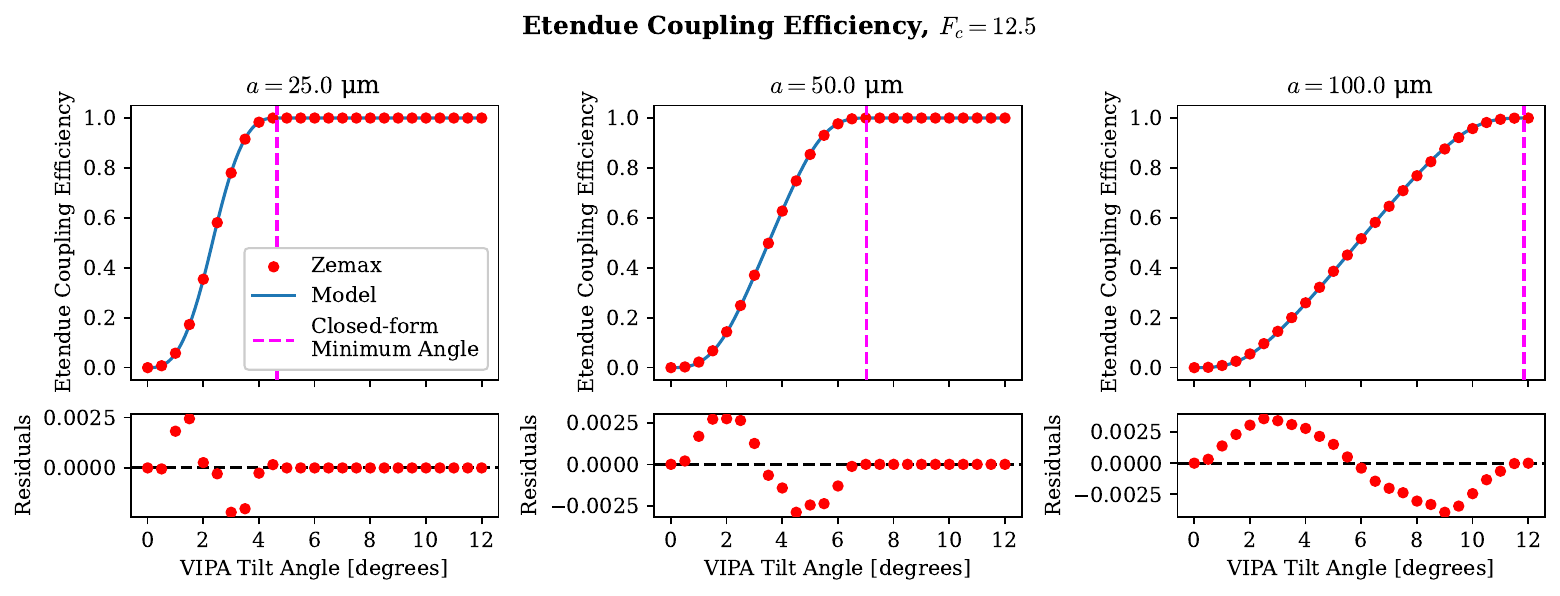}
    \caption{$\eta$ for $F_c=12.5$}
    \label{fig:etendue_eta_Fc12.5}
  \end{subfigure}
  \caption{Etendue coupling efficiency computed from our mathematical model (blue curves, Eq.~(\ref{eq:eta})) and from Zemax non-sequential mode ray-tracing simulations (red points, Eq.~(\ref{eq:eta_zemax})). The magenta vertical dashed lines show the corresponding minimum VIPA tilt angle for no coupling loss, $\tilde{\beta}$, given by Eq.~(\ref{eq:beta_min}). Here, we use $F=5$, $h=2.20$~mm, $n'=1.4494$ (fused silica), $n=1.0003$ (air), $R_1=99.5\%$, $R_2=95\%$, and $L=21$~mm.}
  \label{fig:zemax_etendue_eta}
\end{figure}

\begin{figure}[htbp]
  \centering
  \begin{subfigure}{\textwidth}
    \centering
    \includegraphics[width=\textwidth]{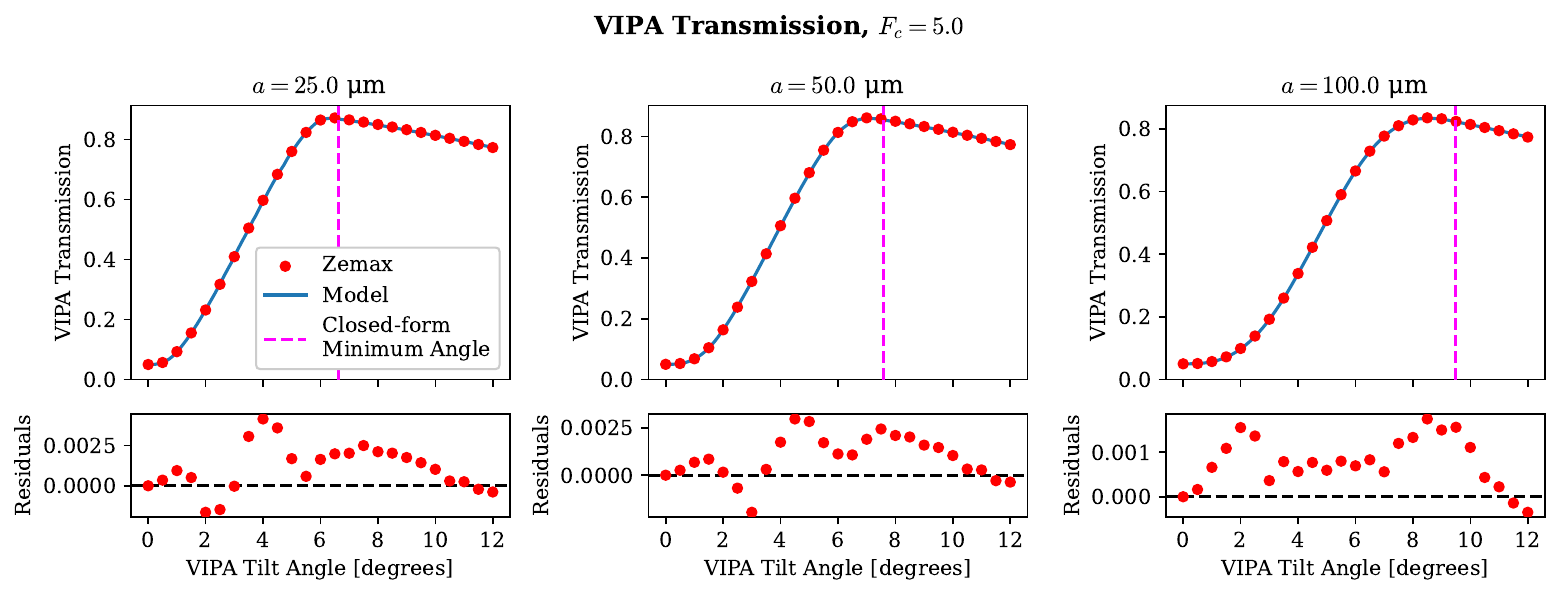}
    \caption{$T_\mathrm{VIPA}$ for $F_c=5$}
    \label{fig:transmission_VIPA_Fc5.0}
  \end{subfigure}
  \hfill
  \begin{subfigure}{\textwidth}
    \centering
    \includegraphics[width=\textwidth]{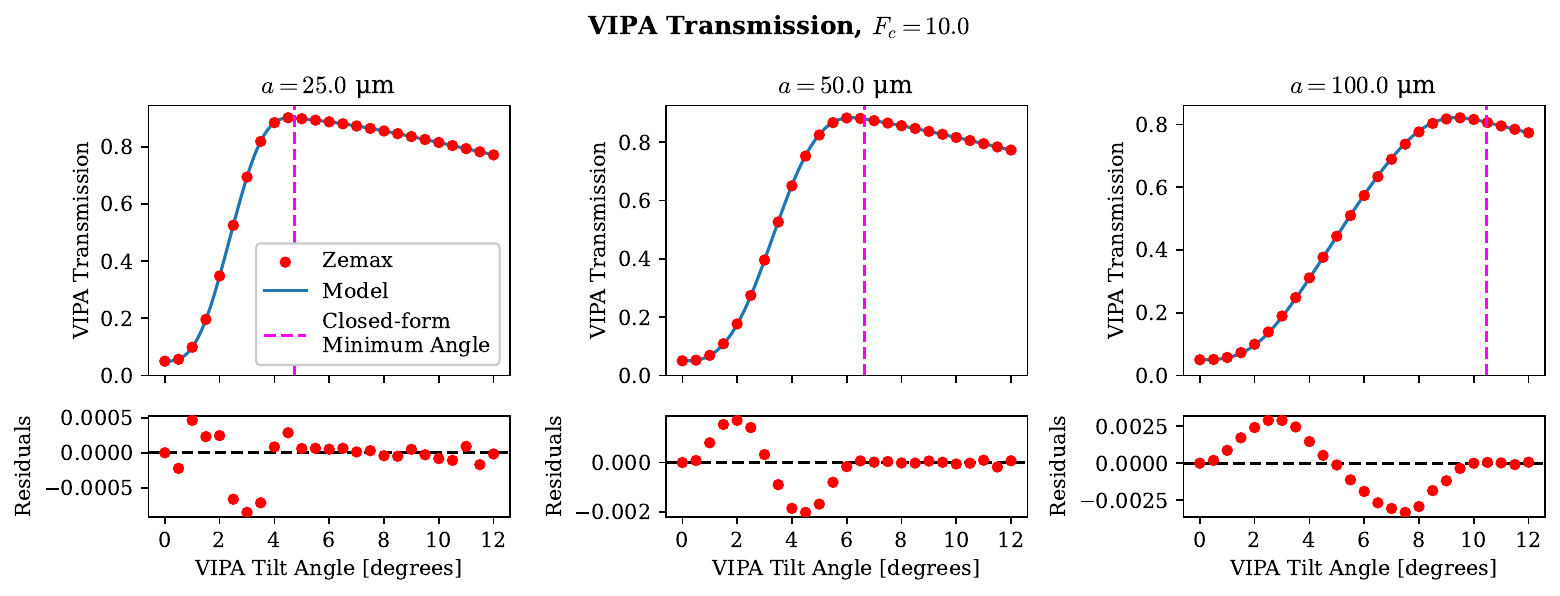}
    \caption{$T_\mathrm{VIPA}$ for $F_c=10$}
    \label{fig:transmission_VIPA_Fc10.0}
  \end{subfigure}
  \hfill
  \begin{subfigure}{\textwidth}
    \centering
    \includegraphics[width=\textwidth]{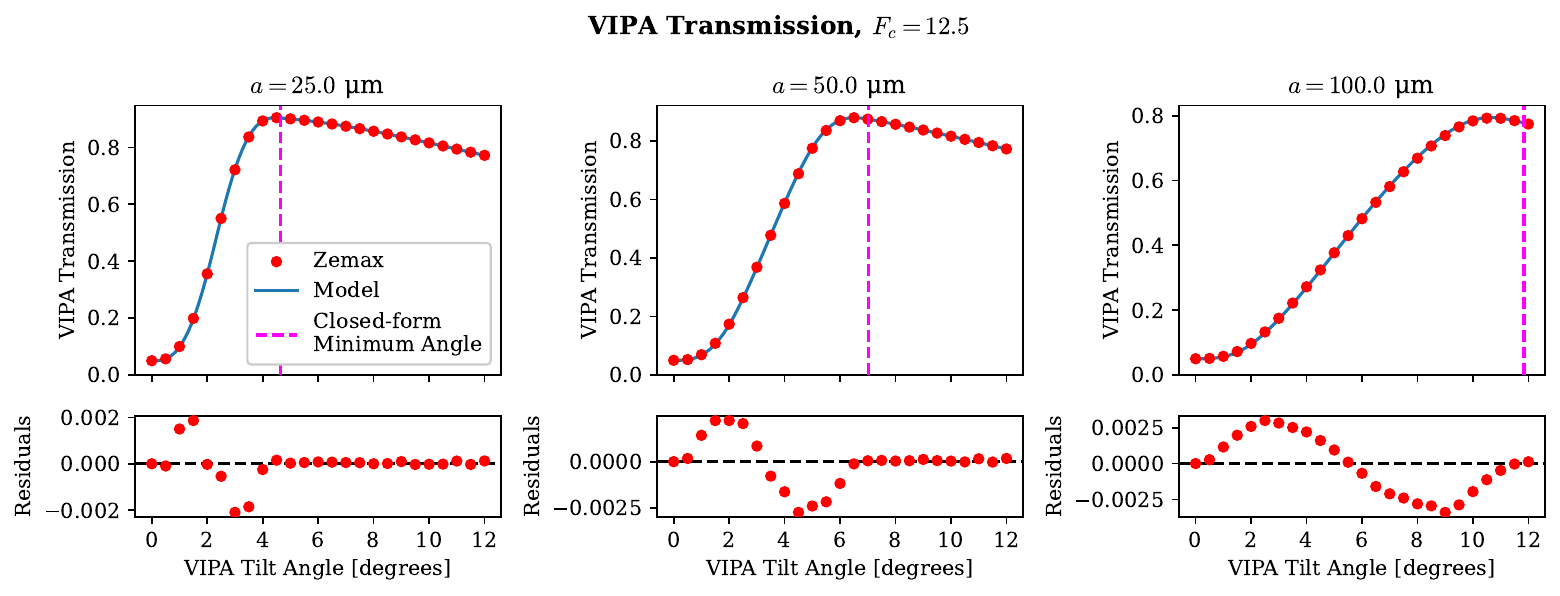}
    \caption{$T_\mathrm{VIPA}$ for $F_c=12.5$}
    \label{fig:transmission_VIPA_Fc12.5}
  \end{subfigure}
  \caption{VIPA transmission computed from our mathematical model (blue curves, Eq.~(\ref{eq:T_VIPA})) and from Zemax non-sequential mode ray-tracing simulations (red points, Eq.~(\ref{eq:T_VIPA_Zemax})). The magenta vertical dashed lines show the corresponding minimum VIPA tilt angle for no coupling loss, $\tilde{\beta}$, given by Eq.~(\ref{eq:beta_min}). Here, we use $F=5$, $h=2.20$~mm, $n'=1.4494$ (fused silica), $n=1.0003$ (air), $R_1=99.5\%$, $R_2=95\%$, and $L=21$~mm.}
  \label{fig:zemax_transmission_VIPA}
\end{figure}


\subsection{Results and Comparison}

Fig.~\ref{fig:zemax_etendue_eta} shows the etendue coupling efficiency $\eta$ as a function of $\beta$, from our mathematical model (Eq.~(\ref{eq:eta})) and Zemax simulations (Eq.~(\ref{eq:eta_zemax})). As seen from the residuals, the Zemax results are in good agreement with our mathematical model of $\eta$. The Zemax results are also consistent with the minimum VIPA tilt angle $\tilde{\beta}$ from Eq.~(\ref{eq:beta_min}), in that when $\beta\geq\tilde{\beta}$, we can see that $\eta=1$ and that the residuals are zero. Hence, etendue is conserved in our model. One limitation to note is that when $\beta<\tilde{\beta}$, there appears to be some structure in the residuals in Fig.~\ref{fig:zemax_etendue_eta}; the residuals appear sinusoidal sometimes. We think that the cause of this is because the ``light source at the VIPA back surface'' description we assumed in Section~\ref{sec:etendue_problem_setup} is an approximation. As a result of the tilted interface between $n$ and $n'$, the best-focus surface of the incoming beam is not simply the plane $N=h$ (there will be coma and other aberrations), and this is not accounted for in our derivation. This is the case even if the optics themselves are coma-free. Nevertheless, all of the residuals of $\eta$ are within $0.005$, which is a small amount.

Fig.~\ref{fig:zemax_transmission_VIPA} shows the VIPA transmission $T_\mathrm{VIPA}$ as a function of $\beta$, from our mathematical model~(Eq.~(\ref{eq:T_VIPA})) and Zemax simulations (Eq.~(\ref{eq:T_VIPA_Zemax})). As seen from the residuals, the Zemax results are in good agreement with our mathematical model of $T_\mathrm{VIPA}$, with most of the residuals of $T_\mathrm{VIPA}$ within $0.0025$. Again, there is some structure in the residuals, which we think can be attributed to the assumptions made in our ``light source at the VIPA back surface'' description. From Fig.~\ref{fig:zemax_transmission_VIPA}, we can see that there are two regimes for $T_\mathrm{VIPA}$. When $\beta$ is small, $T_\mathrm{VIPA}$ is dominated by etendue coupling loss, and so $T_\mathrm{VIPA}$ roughly follows $\eta$. When $\beta$ is large, $T_\mathrm{VIPA}$ is dominated by length-limited loss and monotonically decreases with $\beta$. In this regime, $T_\mathrm{VIPA}$ behaves like the simplified finite geometric series model in Eq.~(\ref{eq:T_VIPA_gs_trunc}). In Fig.~\ref{fig:zemax_transmission_VIPA}, it is interesting to note that the $\beta$ corresponding to the maximum VIPA transmission is not necessarily the same as the minimum VIPA tilt angle for no etendue coupling loss $\tilde{\beta}$. It appears that the maximum VIPA transmission occurs at a slightly shallower tilt angle, as a result of the competition between etendue coupling loss and length-limited loss. Thus, for certain high-throughput applications, one should pick a $\beta$ that is not necessarily $\tilde{\beta}$, but slightly shallower.

Although the etendue coupling efficiency and VIPA transmission can be computed using Zemax non-sequential mode simulations, our mathematical models offer several advantages. First, Zemax is proprietary software. Second, even when Zemax is available, setting up the simulations is quite involved, as detailed in Sections~\ref{sec:Zemax_sim_fiber_and_coupling}, \ref{sec:Zemax_sim_VIPA}, and \ref{sec:Zemax_sim_actual}. In the simulation, the VIPA must be carefully positioned according to the appropriate specific values of $d$ and $T_{0r2}$. Third, evaluating our mathematical models is much faster than running a Zemax non-sequential mode ray trace simulation. For example, the 3D etendue coupled into a VIPA, $G_\mathrm{3D}$~(Eq.~(\ref{eq:3D_etendue_solution})), which is used to compute $\eta$, is a double integral in polar coordinates over the unit disk. This integral can be easily and quickly evaluated numerically using the ``dblquad'' function in the freely-available Python SciPy library \cite{2020SciPy-NMeth_OG}, with one evaluation taking less than a few milliseconds. In comparison, a Zemax non-sequential mode ray trace on the same computer takes three to four orders of magnitude longer to run.

We have not yet validated our mathematical models experimentally, and this is our next step. Nevertheless, as previously mentioned, the Zemax VIPA modeling approach we used, involving two rectangular volumes, was also used by \etalcite{Aryana}{Aryana2026}, and this model has been experimentally validated.


\section{Conclusion}

In this work, we derived a closed-form expression for $\tilde{\beta}$, the minimum tilt angle of a multimode fiber-fed VIPA required for no etendue coupling loss, given by Eq.~(\ref{eq:beta_min}). This expression depends only on 4 parameters: the VIPA thickness $h$, the ratio between the surrounding medium's refractive index and the VIPA's refractive index $n/n'$, the cylindrical lens focal ratio $F_c$, and the input fiber's etendue via the proxy $a/F$.

We also derived a closed-form expression for the 2D etendue that can be coupled into a VIPA, given by Eq.~(\ref{eq:2D_etendue_solution}), and an expression for the 3D etendue that can be coupled into a VIPA, given by Eq.~(\ref{eq:3D_etendue_solution}). Moreover, we derived an expression for the transmitted 3D etendue of a VIPA, given by Eq.~(\ref{eq:G_out}), and hence the transmission of a VIPA, given by Eq.~(\ref{eq:T_VIPA}). Our mathematical models are in agreement with ray-tracing simulations of a VIPA in Zemax OpticStudio non-sequential mode, as presented in Section~\ref{sec:Zemax_sim}.

Our mathematical models establish a clear relationship between the VIPA transmitted etendue and experimentally-relevant physical parameters such as the fiber core radius $a$, the fiber output focal ratio $F$, the cylindrical lens focal ratio $F_c$, and VIPA intrinsic parameters like the VIPA thickness $h$, VIPA HR length $L$, and VIPA refractive index $n'$. The unique dependence of the minimum VIPA tilt angle $\tilde{\beta}$ on the cylindrical lens focal ratio $F_c$ and fiber etendue warrants special attention from users who want to employ VIPAs in optical systems. Our results are useful for the design and optimization of high-throughput optical systems based on multimode fiber-fed VIPAs.




\begin{backmatter}
\bmsection{Funding}
Natural Sciences and Engineering Research Council of Canada.

\bmsection{Acknowledgment}
We thank the anonymous reviewers for their insightful and constructive comments, which helped to improve this work. We thank Kiumars Aryana (NIST) for his helpful guidance on simulating VIPAs in Zemax OpticStudio non-sequantial mode. We thank Tristan McNairn and Calvin Joy (LightMachinery Inc.) for helpful discussions on VIPAs. M.C.H. Leung gratefully acknowledges the support of the Natural Sciences and Engineering Research Council of Canada (NSERC) through an NSERC Postgraduate Scholarship – Doctoral (PGS D).

\bmsection{Disclosures}
The authors declare no competing interests.

\bmsection{Data availability} Data underlying the results presented in this paper are not publicly available at this time but may be obtained from the authors upon reasonable request.

\bmsection{Supplemental document}
See Supplement 1 for supporting content.

\end{backmatter}

\bibliography{bibfile}

\begin{thebibliography}{10}
\newcommand{\enquote}[1]{``#1''}

\bibitem{Shirasaki1996}
M.~Shirasaki, \enquote{{Large angular dispersion by a virtually imaged phased array and its application to a wavelength demultiplexer},} {\protect\JournalTitle{Optics Letters}} \textbf{21}, 366 (1996).

\bibitem{Weiner2012}
A.~M. Weiner, \enquote{Reply to comment on “generalized grating equation for virtually-imaged phased-array spectral dispersions”,} {\protect\JournalTitle{Applied Optics}} \textbf{51}, 8187--8189 (2012).

\bibitem{Xiao2005}
S.~Xiao and A.~M. Weiner, \enquote{{An eight-channel hyperfine wavelength demultiplexer using a virtually imaged phased-array (VIPA)},} {\protect\JournalTitle{IEEE Photonics Technology Letters}} \textbf{17}, 372–374 (2005).

\bibitem{Shirasaki1997}
M.~Shirasaki, \enquote{Chromatic-dispersion compensator using virtually imaged phased array,} {\protect\JournalTitle{IEEE Photonics Technology Letters}} \textbf{9}, 1598--1600 (1997).

\bibitem{Metz2014}
P.~Metz, J.~Adam, M.~Gerken, and B.~Jalali, \enquote{Compact, transmissive two-dimensional spatial disperser design with application in simultaneous endoscopic imaging and laser microsurgery,} {\protect\JournalTitle{Applied Optics}} \textbf{53}, 376--382 (2014).

\bibitem{Li2021}
Z.~Li, Z.~Zang, H.~Y. Fu, \emph{et~al.}, \enquote{{Virtually imaged phased-array-based 2D nonmechanical beam-steering device for FMCW LiDAR},} {\protect\JournalTitle{Applied Optics}} \textbf{60}, 2177--2189 (2021).

\bibitem{NugentGlandorf2012}
L.~Nugent-Glandorf, T.~Neely, F.~Adler, \emph{et~al.}, \enquote{Mid-infrared virtually imaged phased array spectrometer for rapid and broadband trace gas detection,} {\protect\JournalTitle{Optics Letters}} \textbf{37}, 3285--3287 (2012).

\bibitem{Scarcelli2011}
G.~Scarcelli and S.~H. Yun, \enquote{{Multistage VIPA etalons for high-extinction parallel Brillouin spectroscopy},} {\protect\JournalTitle{Optics Express}} \textbf{19}, 10913--10922 (2011).

\bibitem{Bouvet2024}
P.~Bouvet, F.~Clément, A.~Papoz, \emph{et~al.}, \enquote{{Multimode fiber-coupled VIPA spectrometer for high-throughput Brillouin imaging of biological samples},} {\protect\JournalTitle{Journal of Physics: Photonics}} \textbf{6}, 025010 (2024).

\bibitem{Bourdarot2018}
G.~Bourdarot, E.~Le~Coarer, D.~Mouillet, \emph{et~al.}, \enquote{{Experimental test of a 40 cm-long R=100 000 spectrometer for exoplanet characterisation},} in \emph{Ground-based and Airborne Instrumentation for Astronomy VII,}  vol. 10702 of \emph{Proc. SPIE} (2018), p. 107025Y.

\bibitem{Zhu2020}
X.~Zhu, D.~Lin, Z.~Hao, \emph{et~al.}, \enquote{{A VIPA Spectrograph with Ultra-high Resolution and Wavelength Calibration for Astronomical Applications},} {\protect\JournalTitle{The Astronomical Journal}} \textbf{160}, 135 (2020).

\bibitem{Carlotti2022}
A.~Carlotti, A.~Bidot, D.~Mouillet, \emph{et~al.}, \enquote{{On-sky demonstration at Palomar Observatory of the near-IR, high-resolution VIPA spectrometer},} in \emph{Ground-based and Airborne Instrumentation for Astronomy IX,}  vol. 12184 of \emph{Proc. SPIE} (2022), p. 121841I.

\bibitem{Zhu2023}
X.~Zhu, D.~Lin, Z.~Zhang, \emph{et~al.}, \enquote{{Dispersion Characteristics of the Multi-mode Fiber-fed VIPA Spectrograph},} {\protect\JournalTitle{The Astronomical Journal}} \textbf{165}, 228 (2023).

\bibitem{Leung2025}
M.~C.~H. Leung, D.~Charbonneau, A.~Szentgyorgyi, \emph{et~al.}, \enquote{{VIPER: a high-resolution multimode fiber-fed VIPA spectrograph concept for characterizing exoplanet atmospheric escape},} in \emph{Techniques and Instrumentation for Detection of Exoplanets XII,}  vol. 13627 of \emph{Proc. SPIE} (2025), p. 136271P.

\bibitem{SalehTeich}
B.~E.~A. Saleh and M.~C. Teich, \emph{{Fundamentals of Photonics}} (John Wiley \& Sons, Inc., 1991). {Chap. 8}.

\bibitem{nonimagingoptics_chaves}
J.~Chaves, \emph{{Introduction to Nonimaging Optics}} (CRC, Taylor \& Francis Group, 2016), 2nd ed. {Chap. 4, 18}.

\bibitem{Hu2015}
X.~Hu, Q.~Sun, J.~Li, \emph{et~al.}, \enquote{Spectral dispersion modeling of virtually imaged phased array by using angular spectrum of plane waves,} {\protect\JournalTitle{Optics Express}} \textbf{23}, 1--12 (2015).

\bibitem{Winston2018}
R.~Winston, L.~Jiang, and M.~Ricketts, \enquote{Nonimaging optics: a tutorial,} {\protect\JournalTitle{Advances in Optics and Photonics}} \textbf{10}, 484 (2018).

\bibitem{Pedroti_Optics}
F.~L. Pedrotti, L.~M. Pedrotti, and L.~S. Pedrotti, \emph{{Introduction to Optics}} (Cambridge University, 2017), 2nd ed. {Chap. 20}.

\bibitem{Brooker_Optics}
G.~Brooker, \emph{{Modern Classical Optics}} (Oxford University, 2003). {Chap. 11}.

\bibitem{Miller2012}
D.~A.~B. Miller, \enquote{All linear optical devices are mode converters,} {\protect\JournalTitle{Optics Express}} \textbf{20}, 23985--23993 (2012).

\bibitem{Miller2019}
D.~A.~B. Miller, \enquote{Waves, modes, communications, and optics: a tutorial,} {\protect\JournalTitle{Advances in Optics and Photonics}} \textbf{11}, 679--825 (2019).

\bibitem{Piestun2000}
R.~Piestun and D.~A.~B. Miller, \enquote{Electromagnetic degrees of freedom of an optical system,} {\protect\JournalTitle{Journal of the Optical Society of America A}} \textbf{17}, 892--902 (2000).

\bibitem{Epworth1979}
R.~E. Epworth, \enquote{{Phenomenon of modal noise in fiber systems},} in \emph{Optical Fiber Communication,}  (OSA, 1979), p. ThD1.

\bibitem{Baudrand2001}
J.~Baudrand and G.~A.~H. Walker, \enquote{{Modal Noise in High‐Resolution, Fiber‐fed Spectra: A Study and Simple Cure},} {\protect\JournalTitle{Publications of the Astronomical Society of the Pacific}} \textbf{113}, 851--858 (2001).

\bibitem{Lemke2011}
U.~Lemke, J.~Corbett, J.~Allington-Smith, and G.~Murray, \enquote{{Modal noise prediction in fibre spectroscopy - I. Visibility and the coherent model: Modal noise prediction in fibre spectroscopy - I},} {\protect\JournalTitle{Monthly Notices of the Royal Astronomical Society}} \textbf{417}, 689--697 (2011).

\bibitem{Leung2025MS}
M.~C.~H. Leung, C.~Jurgenson, A.~Szentgyorgyi, \emph{et~al.}, \enquote{{Crank-rocker optical fiber mode scrambler prototype for the GMT-Consortium Large Earth Finder (G-CLEF)},} in \emph{Techniques and Instrumentation for Detection of Exoplanets XII,}  vol. 13627 of \emph{Proc. SPIE} (2025), p. 136271J.

\bibitem{Aryana2026}
K.~Aryana, D.~M. Bailey, S.~I. Woods, and A.~J. Fleisher, \enquote{{Bridging Theory and Experiment in Virtually Imaged Phased Array (VIPA) Spectrometers},} {\hypersetup{hidelinks}\url{https://arxiv.org/abs/2601.08589}}, DOI: 10.48550/ARXIV.2601.08589 (2026).

\bibitem{vanVught2024_ZOSPy}
L.~van Vught, C.~Haasjes, and J.-W.~M. Beenakker, \enquote{{ZOSPy: optical ray tracing in Python through OpticStudio},} {\protect\JournalTitle{Journal of Open Source Software}} \textbf{9}, 5756 (2024).

\bibitem{2020SciPy-NMeth_OG}
P.~Virtanen, R.~Gommers, T.~E. Oliphant, \emph{et~al.}, \enquote{{{SciPy} 1.0: Fundamental Algorithms for Scientific Computing in Python},} {\protect\JournalTitle{Nature Methods}} \textbf{17}, 261--272 (2020).

\end{thebibliography}

\end{document}